\documentclass[11pt,a4paper]{article}

\usepackage[margin=1in]{geometry}
\usepackage{amsmath,amssymb,amsfonts,mathtools}
\usepackage{bm}
\usepackage{booktabs}
\usepackage{array}
\usepackage{enumitem}
\usepackage[table]{xcolor}
\usepackage{hyperref}
\usepackage{cite}
\usepackage[nameinlink,noabbrev]{cleveref}
\usepackage{bbm}
\usepackage{graphicx}

\hypersetup{
    colorlinks=true,
    linkcolor=blue!60!black,
    citecolor=blue!60!black,
    urlcolor=blue!60!black
}

\newcommand{\bb}[1]{\mathbb{#1}}
\newcommand{\Cl}[1]{\bb{C}\ell(#1)}
\newcommand{\Mat}{\mathrm{Mat}}
\newcommand{\Aut}{\mathrm{Aut}}

\newcommand{\SU}{\mathrm{SU}}
\newcommand{\U}{\mathrm{U}}
\newcommand{\SO}{\mathrm{SO}}
\newcommand{\Spin}{\mathrm{Spin}}

\newcommand{\Gen}[1]{\mathcal{S}_{#1}}
\newcommand{\ii}{\mathrm{i}}

\newcommand{\su}{\mathfrak{su}}
\newcommand{\spin}{\mathfrak{spin}}

\newcommand{\SUthreeC}{\SU(3)_C}
\newcommand{\SUtwoL}{\SU(2)_L}
\newcommand{\UoneY}{\U(1)_Y}
\newcommand{\Uoneem}{\U(1)_{\mathrm{em}}}

\title{Sedenions, Clifford Algebras, and Three Fermion Generations:\newline A Focused Review}

\author{Niels Gresnigt\thanks{\protect\raggedright Department of Physics, Xi'an Jiaotong-Liverpool University, Suzhou, P.R. China.\\ Email: \href{mailto:niels.gresnigt@xjtlu.edu.cn}{\texttt{niels.gresnigt@xjtlu.edu.cn}}}}

\date{}

\begin{document}

\maketitle

\begin{abstract}

The existence of three fermion generations remains one of the unexplained structural features of the Standard Model. This article provides a focused review of an algebraic framework developed in a series of recent works, in which an intrinsic $S_3$ family symmetry relates three gauge-equivalent fermion sectors.

The framework is motivated by division algebra and Clifford algebra constructions in which the complex octonions and $\bb{C}\ell(6)$ organise the colour and electromagnetic quantum numbers of a single generation. Continuing the Cayley--Dickson sequence, the sedenions provide an algebraic setting in which the automorphism structure contains an intrinsic $S_3$ factor, while their complexified left-multiplication operators generate an associative endomorphism algebra isomorphic to $\Cl{8}$. In the $\Cl{8}$ formulation, the order-three family action produces three linearly independent fermion sectors while leaving the $\mathfrak{su}(3)_C\oplus\mathfrak{u}(1)_{\rm em}$ gauge algebra untriplicated. Its extension to $\bb{C}\ell(10)$ incorporates a single $\mathfrak{su}(3)_C\oplus\mathfrak{su}(2)_L\oplus\mathfrak{u}(1)_Y$ gauge algebra, whose generators are invariant under the family action, and each generation includes a right-handed neutrino that is sterile under the SM gauge interactions.

We place this construction in context through a focused comparison with representative algebraic and family-symmetry approaches to three generations, including triality-based proposals, and clarify why the family symmetry used here is distinct from standard $\Spin(8)$ triality despite the common appearance of $\Cl{8}$ and $S_3$. The framework remains algebraic and representation theoretic rather than a complete dynamical theory; family-symmetry breaking, realistic fermion masses and mixing, and a dynamical account of the gauge and matter sectors remain open problems.

\end{abstract}

\vspace{0.5em}


\vspace{1em}

\tableofcontents

\section{Introduction}
\label{sec:introduction}

The Standard Model (SM) provides an extraordinarily successful description of the known elementary particles and their gauge interactions, yet it leaves several structural features unexplained. These include the origin of the gauge group, the pattern of fermion representations and charge assignments, and the existence of three observed fermion generations. Algebraic approaches to particle physics ask whether some of these features can be understood from the representation theory and internal structure of distinguished mathematical systems, including division algebras, Clifford algebras, exceptional Jordan and Lie algebras, and related topological constructions.

In this review, algebraic unification is understood in a broad sense. It refers to an approach in which features of the SM are organised, constrained, or partially explained by the internal structure of a chosen algebraic system. This need not involve embedding the SM gauge group into a single simple Lie group, as in a conventional grand unified theory, nor does it imply that a complete dynamical theory has already been obtained. The framework considered here should be understood in this sense: it provides an algebraic construction of fermion generations and gauge structure, rather than a dynamical unification of all interactions.

The main focus of this article is a sedenion-motivated framework for three generations of SM fermions based on the complex Clifford algebras $\Cl{8}$ and $\Cl{10}$. We reconstruct the development of this framework in a common notation, beginning with its motivation from the Cayley--Dickson algebra of sedenions $\bb{S}$. The complexified left-multiplication operators of the sedenions generate an associative endomorphism algebra isomorphic to $\Cl{8}$, while the intrinsic $S_3$ structure associated with the automorphisms of $\bb{S}$ is used to construct a family action on selected Clifford spinors. The framework is then extended to $\Cl{10}$ to incorporate the full SM gauge algebra.

The review is timely because these ingredients have now been brought together within a common algebraic setting. In the $\Cl{8}$ formulation, the order-three family action generates three linearly independent, gauge-equivalent fermion sectors, while the colour and electromagnetic generators form a single $\mathfrak{su}(3)_C\oplus\mathfrak{u}(1)_{\rm em}$ gauge algebra. The $\Cl{10}$ extension incorporates weak isospin and hypercharge, giving a common $\mathfrak{su}(3)_C\oplus\mathfrak{su}(2)_L\oplus\mathfrak{u}(1)_Y$ gauge algebra whose generators are invariant under the $S_3$ family action. The fermion sectors, gauge generators, and family symmetry are therefore represented together within the same associative Clifford algebra. At the same time, the construction remains algebraic and representation theoretic: it does not yet determine family symmetry breaking, realistic fermion masses and mixing, or the dynamics of the gauge and matter sectors.

A central purpose of the review is to explain how this formulation emerged. The earliest sedenion model attempted to associate the three generations directly with three overlapping octonion subalgebras of $\bb{S}$. Although this revealed a natural threefold structure, it did not naturally provide independent generation spaces with a single common gauge action. The later construction shifted the emphasis from the subalgebras themselves to an $S_3$-generated orbit of fermion subspaces within $\Cl{8}$. The mature formulation is therefore not merely the earlier model rewritten in a different Clifford basis, but represents a change in the underlying generation mechanism.

Alongside this reconstruction, the review has two broader aims. The first is to trace the historical development of the underlying ideas, including the representation of spinors by minimal ideals of Clifford algebras, fermionic oscillator constructions, the use of division and Clifford algebras in particle models, triality, and discrete family symmetries. The second is to compare the resulting generation mechanism with representative alternatives based on division and Clifford algebras, $\Spin(8)$ triality, exceptional Jordan algebras, dimensional reduction, and conventional $S_3$ flavour symmetry. These comparisons are intended to distinguish the different mathematical meanings assigned to a threefold structure, rather than to suggest that the various approaches have identical aims.

\subsection{Scope and aims of the review}

This paper is intended to be focused but largely self-contained. It provides sufficient background on Clifford algebras, minimal ideals, the Cayley--Dickson construction, sedenion automorphisms, and the use of $S_3$ as a family symmetry to make the main arguments accessible without requiring continual reference to earlier papers. Detailed calculations are included where they are needed to establish the construction, but the review does not reproduce every basis-level computation.

The primary subject remains the sedenion-motivated route and the development of its $\Cl{8}$ and $\Cl{10}$ formulations. The historical discussion is therefore selective, concentrating on ideas that lead directly to the present construction or clarify its mathematical setting. Likewise, the comparative discussion is not intended as a comprehensive survey of algebraic particle physics. It considers representative proposals that illuminate different possible origins of three generations and provide a useful basis for assessing the scope of the present framework.

The review addresses four connected sets of questions. First, what is gained by continuing the Cayley--Dickson sequence from the octonions to the sedenions, despite the fact that the sedenions are no longer division algebras, and how do their complex left multiplication maps generate $\Cl{8}$? Second, how are the relevant fermion subspaces constructed from primitive idempotents, minimal ideals, and semi-spinors, and how is the sedenion-induced $S_3$ family action realised within the same Clifford algebra? Third, how does this action generate three independent fermion sectors with a common electrocolour symmetry, and how does the extension to $\Cl{10}$ incorporate the full SM gauge algebra and its corresponding fermion representations? Finally, how did these ideas develop historically, and how does the resulting mechanism differ from other algebraic and family-symmetry approaches to three generations?

\subsection{Notation}

The notation used throughout the review is summarised in Table~\ref{tab:notation-conventions}. Definitions and conventions specific to individual constructions are introduced where they first become relevant.

\begin{table}[htbp]
\centering
\small
\begin{tabular}{p{0.23\textwidth}p{0.67\textwidth}}
\toprule
Notation & Meaning \\
\midrule
$\bb{R}$ & The real numbers. \\
$\bb{C}$ & The complex numbers; also used as the scalar field for complexified algebras. \\
$\bb{H}$ & The quaternions. \\
$\bb{O}$ & The octonions. \\
$\bb{S}$ & The real algebra of sedenions. \\
$\bb{T}$ & Dixon's algebra $\bb{R}\otimes\bb{C}\otimes\bb{H}\otimes\bb{O}$, the tensor product of the four normed division algebras. \\
$J_3(\bb{O})$ & The exceptional Jordan algebra of $3\times3$ Hermitian matrices over $\bb{O}$, used in exceptional-algebra approaches to SM structure and the generation problem. \\
$\bb{C}\otimes\bb{O}$ & The complexified octonions, whose composed left-multiplication maps generate $\operatorname{End}_{\bb{C}}(\bb{C}\otimes\bb{O})\simeq\Cl{6}$ in the one-generation constructions reviewed here. \\
$\bb{C}\otimes\bb{S}$ & The complexified sedenions, whose composed left-multiplication maps generate $\operatorname{End}_{\bb{C}}(\bb{C}\otimes\bb{S})\simeq\Cl{8}$ in the present construction. \\
$s_i$ & Sedenion basis elements, with $s_0=1$ and $s_i$ imaginary for $i=1,\ldots,15$. \\
$e_i$ & Clifford generators or left-action matrices, distinct from the sedenion basis elements $s_i$. \\
$\Cl{n}$ & Complex Clifford algebra on $n$ generators. \\
$\Cl{6}$ & Clifford algebra associated with one-generation constructions based on $\bb{C}\otimes\bb{O}$. \\
$\Cl{8}$ & Clifford algebra generated by complex left actions associated with $\bb{C}\otimes\bb{S}$. \\
$\Cl{10}$ & Electroweak extension of the $\Cl{8}$ construction. \\
$e_{i_1\cdots i_k}$ & Ordered Clifford product $e_{i_1}\cdots e_{i_k}$. \\
$\omega_8$ & Pseudoscalar $e_1e_2\cdots e_8$ of $\Cl{8}$. \\
$a_i,a_i^\dagger$ & Witt-basis annihilation and creation operators. \\
$f_{\epsilon_1\epsilon_2\cdots}$ & Primitive idempotent built from choices of $a_i a_i^\dagger$ or $a_i^\dagger a_i$. \\
$I_i$ & Minimal left ideal labelled by the corresponding primitive idempotent or matrix unit. \\
$I_i^\pm$ & Semi-spinor subspaces of a minimal ideal. \\
$\psi_3$ & Order-three generator of the $S_3$ family action. \\
$\epsilon$ & Order-two generator of the $S_3$ family action. \\
$\Gen{1},\Gen{2},\Gen{3}$ & Fermion-generation sectors, not to be confused with the sedenion algebra $\bb{S}$. \\
$\Lambda_a$ & Colour generators, $a=1,\ldots,8$. \\
$Q$ & $S_3$-invariant electromagnetic charge operator in the $\Cl{8}$ construction. \\
$Q'$ & Physical electromagnetic charge operator in the $\Cl{10}$ electroweak extension. \\
$T_i$ & Weak $\SUtwoL$ generators. \\
$Y$ & Hypercharge, defined by $Y=2(Q'-T_3)$. \\
\bottomrule
\end{tabular}
\caption{Notation and conventions used throughout the review.}
\label{tab:notation-conventions}
\end{table}

\subsection{Organisation of the review}

Section~\ref{sec:historical-remarks} places the framework in its historical and conceptual context. It reviews the representation of spinors by Clifford ideals, early constructions based on division algebras and fermionic oscillators, one-generation models, topological representations of algebraic spinors, dimensional reduction, triality, exceptional structures, and the development from overlapping octonion subalgebras of the sedenions to an $S_3$-generated family structure. Section~\ref{sec:algebraic-foundations} then develops the algebraic foundations used in the remainder of the paper. It reviews the $\Cl{6}$ one-generation construction, introduces the Cayley--Dickson construction and the sedenions, and derives the intrinsic $S_3$ automorphisms, the complex left-action algebra $\Cl{8}$, and the associated Witt-basis and minimal-ideal structures.

Section~\ref{sec:S3-in-Cl8} constructs the sedenion-induced family action within $\Cl{8}$ and describes its action on the Clifford generators, Witt bases, primitive idempotents, and spinor subspaces. Section~\ref{sec:three-generations-Cl8} constructs the three fermion sectors and their common action under the $\mathfrak{su}(3)_C\oplus\mathfrak{u}(1)_{\rm em}$ gauge algebra. Section~\ref{sec:Cl10-electroweak-extension} extends the framework to $\Cl{10}$, where these $S_3$-related sectors transform under a common $\mathfrak{su}(3)_C\oplus\mathfrak{su}(2)_L\oplus\mathfrak{u}(1)_Y$ gauge algebra.

Section~\ref{sec:comparative-analysis} compares this generation mechanism with representative division algebra and Clifford algebra constructions, triality-based models, exceptional Jordan algebra approaches, octonion dimensional reduction, and conventional $S_3$ family-symmetry models. Section~\ref{sec:conclusions-outlook} summarises the established results and discusses the principal limitations and open problems. The sedenion multiplication convention used throughout the paper is recorded in Appendix~\ref{app:sedenion-multiplication-table}.

\section{Historical context and related algebraic approaches}
\label{sec:historical-remarks}

The historical background relevant to this framework brings together several ideas: the representation of spinors by minimal Clifford ideals, the use of octonions and $\Aut(\bb{O})=G_2$ in the description of colour, oscillator constructions based on $\Cl{6}$ and $\Cl{10}$, and later work involving $\Cl{8}$, triality, exceptional Jordan algebras, and discrete family symmetries. The main focus of this review is a sedenion-motivated route developed in a series of recent works \cite{Gillard2019,Gresnigt2023,gourlay2024algebraic,Gresnigt2026PLB}, in which the associative algebra generated by complexified sedenion left-multiplication operators is isomorphic to $\Cl{8}$ and the intrinsic $S_3$ structure of the sedenions is used to construct a family action, followed by the extension to $\Cl{10}$. The present section places this development in its historical context, while Section~\ref{sec:comparative-analysis} compares the resulting construction with representative alternative algebraic proposals for three fermion generations. For broader background on nonassociative algebras in physics, see Lõhmus, Paal, and Sorgsepp \cite{Lohmus1994}.

\subsection{Algebraic spinors and minimal ideals}

The realisation of spinors within Clifford algebras has a long history. Early work by Juvet and Sauter connected Dirac operators and Maxwell equations with Clifford structures \cite{Juvet1930,Sauter1930}, while Riesz developed an influential treatment of spinors in terms of Clifford ideals \cite{riesz2013clifford}. The point relevant to the present review is that a spinor space may be realised as a minimal left ideal of a Clifford algebra. In many of the constructions reviewed below, such ideals, or selected semi-spinor subspaces within them, are identified with fermionic state spaces.

A standard construction starts from a Witt basis of nilpotent creation and annihilation operators. Products of these operators define primitive idempotents, while the creation operators generate an exterior-algebra basis for the corresponding minimal left ideals \cite{Ablamowicz1995}. This construction is reviewed for the $\Cl{6}$ one-generation construction in Section~\ref{sec:Cl6-one-generation-template} and developed explicitly for $\Cl{8}$ in Section~\ref{subsec:Cl8-witt-minimal-ideals}. In the present framework, selected semi-spinor subspaces of minimal ideals in $\Cl{8}$ and $\Cl{10}$ provide the fermion sectors, while the gauge generators and $S_3$ family action are represented within the same associative Clifford algebra, acting through commutators and automorphisms, respectively.

\subsection{Octonions, colour, and early fermionic oscillator models}

The connection between the highest-dimensional normed division algebra, the octonions $\bb{O}$, and quark colour was developed in early work by G\"{u}naydin and G\"{u}rsey, who related the resulting $\SU(3)$ triplet structure to quark colour \cite{Gunaydin1973}.\footnote{The basis used by G\"{u}naydin and G\"{u}rsey is obtained by adjoining the ordinary complex unit to the real octonions and forming idempotent and nilpotent combinations of octonion basis elements. Their construction is therefore naturally understood inside $\bb{C}\otimes\bb{O}$ rather than within the real division algebra $\bb{O}$ alone. The latter contains no nontrivial idempotents or nilpotent elements, whereas $\bb{C}\otimes\bb{O}$ admits the projector and nilpotent structures used in the colour decomposition.} The underlying group-theoretic observation is that
\begin{eqnarray}
\Aut(\bb{O})&=&G_2,\qquad \SU(3)\simeq\operatorname{Stab}_{G_2}(u),
\end{eqnarray}
where $u$ is a chosen imaginary octonion unit. The stabiliser subgroup therefore provides a natural algebraic origin for colour symmetry.

A parallel line of development used quantised Grassmann variables, or equivalently fermionic oscillators, to describe the internal degrees of freedom of quarks and leptons. Barducci et al.\ introduced three oscillator pairs $a_i,a_i^\dagger$, $i=1,2,3$, transforming as colour triplets \cite{Barducci1977}. In the notation used here, three such pairs generate the complex Clifford algebra $\Cl{6}$. The same Clifford algebra is obtained as an associative endomorphism algebra of $\bb{C}\otimes\bb{O}$. In Dixon’s treatment, the composed left- and right-multiplication maps each generate this algebra, whereas Furey’s later one-generation construction uses composed left-multiplication maps \cite{Dixon1990,Dixon1994,furey2016standard}.

Casalbuoni and Gatto developed the oscillator approach into a unified description of quarks and leptons \cite{Casalbuoni1979,Casalbuoni1980}. Additional colour-singlet oscillators were introduced to incorporate the remaining internal degrees of freedom. Four oscillator pairs give a $16$-dimensional Fock space capable of accommodating the states of one fermion generation, while five pairs generate $\Cl{10}$ and allow a bilinear weak $\SU(2)$ action. These works also explored family replication by adjoining further colour-singlet oscillator pairs. With $m$ additional pairs, the Fock space decomposes into $2^m$ family copies, so this mechanism produces replication in powers of two rather than selecting exactly three families. These early models established the oscillator language later reformulated in terms of Witt bases, primitive idempotents, and minimal ideals. The corresponding $\Cl{6}$ one-generation construction is reviewed explicitly in Section~\ref{sec:Cl6-one-generation-template}, while alternative algebraic mechanisms for family replication are compared in Section~\ref{sec:comparative-analysis}.

\subsection{The Dixon algebra and division algebra unification}
\label{sec:dixon-algebra}

Dixon developed a systematic algebraic framework based on the tensor product of the four real normed division algebras
\begin{eqnarray}
\bb{T}&=&\bb{R}\otimes\bb{C}\otimes\bb{H}\otimes\bb{O}.
\end{eqnarray}
The guiding idea is that the division algebras should not be regarded merely as convenient coefficient systems, but as distinguished mathematical structures whose exceptional properties are reflected in the algebraic organisation of the SM \cite{Dixon1990,Dixon1994,Dixon2010,dixon2014}. The factors of $\bb{T}$ play different roles: $\bb{C}$ supplies phase and projector structures, $\bb{H}$ is associated with spin and weak isospin, and $\bb{O}$ provides the colour-related structure arising from $G_2$ and its $\SU(3)$ stabiliser subgroups.

A central feature of Dixon's approach is that division algebras and their tensor products are treated as spinor, or hyperspinor, spaces in their own right. Their left and right multiplication maps, defined by $L_a[z]=az$ and $R_a[z]=za$, generate associative algebras of linear transformations. This remains true when the underlying algebra is nonassociative, since compositions of multiplication maps are ordinary linear maps and are associative under composition. For the octonions, the algebras generated separately by the left and right actions are each isomorphic to the real Clifford algebra $C\ell(0,6)$, while complexification gives
\begin{eqnarray}
\bb{O}_L\simeq\bb{O}_R&\simeq&\Mat(8,\bb{R})\simeq C\ell(0,6),\\
(\bb{C}\otimes\bb{O})_L&\simeq&\Mat(8,\bb{C})\simeq\Cl{6}.
\end{eqnarray}
Thus $\bb{O}$ and $\bb{C}\otimes\bb{O}$ may be regarded as spinor modules for their respective multiplication algebras. In Dixon's formulation, the spinors belong to the underlying division-algebraic space, whereas the Clifford algebra acts on that space through multiplication maps. This is conceptually distinct from identifying spinors directly with minimal left ideals inside the Clifford algebra. The relation between multiplication algebras and Clifford algebras is developed more explicitly in Section~\ref{sec:multiplication-algebras}.

Within $\bb{T}$, the octonion factor supplies the colour structure described in the preceding subsection. Idempotent projectors select a $\U(3)$ symmetry and organise the relevant states into the characteristic representation
\begin{eqnarray}
\mathbf{1}\oplus\mathbf{3}\oplus\overline{\mathbf{3}}\oplus\mathbf{1},
\end{eqnarray}
corresponding to leptons, quarks, antiquarks, and antileptons. The quaternion factor in $\bb{T}$ plays a complementary role. Its left and right multiplication algebras commute, so right multiplication by unit quaternions provides an internal $\SU(2)$ action relative to the left action algebra. Dixon interprets this symmetry as the source of weak isospin. Together with the $\U(1)$ structure associated with the complex factor, these ingredients organise one family of leptons and quarks together with its antifamily under a $\U(1)\times\SU(2)\times\SU(3)$ symmetry.

Dixon also proposed a separate construction intended to account for family replication \cite{Dixon2004}. Beginning with a one-family hyperspinor space based on $\bb{T}^2$, he introduced the enlarged structure
\begin{eqnarray}
\bb{T}^6&=&\bb{C}^{1}\otimes\bb{H}^{2}\otimes\bb{O}^{3}.
\end{eqnarray}
This provides a structured three-family extension of the Dixon algebra, although the full multiplication algebra associated with $\bb{T}^6$ is not itself a Clifford algebra.

Dixon’s work provides important historical and conceptual background for the sedenion-motivated construction reviewed here, particularly through the distinction between a nonassociative algebraic space and the associative multiplication algebra acting upon it. The sedenion-motivated construction is not obtained by simply replacing the octonion factor $\bb{O}$ in $\bb{T}$ with $\bb{S}$. Instead, the sedenions provide a route to $\Cl{8}$ together with the $S_3$ structure from which the family action is constructed, while the fermion states are represented within the resulting associative Clifford algebra. The historical development of this construction is described in Section~\ref{sec:historical-development}, and its relation to alternative three-generation mechanisms is considered in Section~\ref{sec:comparative-analysis}.

\subsection{Division algebras, ladder operators, and one-generation models}
\label{sec:division-algebraic-ladder-operators}

A later formulation was developed by Furey, building on the earlier fermionic-oscillator constructions and on the use of complex octonions in Dixon's work \cite{furey2016standard}. Three ladder-operator pairs obtained from $\bb{C}\otimes\bb{O}$ generate the complex Clifford algebra $\Cl{6}$ through their composed left actions. Two conjugate minimal left ideals are selected to represent the particle and antiparticle states of one SM generation. On these ideals, the commutator action of $\mathfrak{su}(3)_C$ reduces to a one-sided left action and, together with the corresponding $\mathfrak{u}(1)_{\rm em}$ generator, reproduces the unbroken $\SU(3)_C\times\U(1)_{\rm em}$ quantum numbers. Furey's construction thus brought the earlier oscillator and division-algebra approaches together in a compact formulation based on ladder operators, primitive idempotents, and minimal ideals. The construction is reviewed explicitly in Section~\ref{sec:Cl6-one-generation-template}.

Furey subsequently combined ladder operators arising from the quaternion and octonion sectors to form five pairs generating $\Cl{10}$, and used their symmetry to recover the full SM gauge algebra \cite{Furey2018a}. Requiring transformations to preserve the distinction between the underlying algebraic actions restricts the larger $\SU(5)$ ladder symmetry to
\begin{eqnarray}
\big(\SU(3)_C\times\SU(2)_L\times\U(1)_Y\big)/\bb{Z}_6.
\end{eqnarray}
This provides a division algebra based description of the gauge structure and representation content of one fermion generation.

Stoica developed a related ideal-based construction with a different starting point \cite{stoica2018leptons}. Rather than deriving $\Cl{6}$ from $\bb{C}\otimes\bb{O}$, he begins with a complex three-dimensional Hermitian space $\chi$ and its dual, giving
\begin{eqnarray}
\Cl{\chi^\dagger\oplus\chi}\simeq\Cl{6}.
\end{eqnarray}
All eight minimal left ideals are then used to represent the states of one SM family. Opposite-sided actions distinguish the Lorentz and weak structures from the colour and electromagnetic symmetries. Like Furey's construction, this is a one-generation representation-building framework rather than an intrinsic mechanism for family replication.

A further related construction was developed in \cite{Gresnigt2020StandardModel}. Two minimal left ideals of $\Cl{6}$ carrying the electrocolour quantum numbers were combined with two minimal right ideals of $\Cl{4}$ carrying the chiral weak structure. The resulting fermion states can be embedded into minimal left ideals of $\Cl{10}$ while retaining the separate $\Cl{6}$ and $\Cl{4}$ structures associated with the physical states. This gives the fermion content and full $\SUthreeC\times\SUtwoL\times\UoneY$ gauge symmetries of one generation, but does not itself provide a mechanism for family replication.

The principal conceptual distinction from Dixon's approach concerns the location of the fermion states. Dixon treats the underlying tensor product of division algebras as a hyperspinor module on which the associated multiplication algebra acts, whereas the later ideal-based constructions represent fermions by minimal ideals within the associated Clifford algebra. These viewpoints are closely related, but they assign different roles to the underlying algebraic space and its multiplication algebra. The ideal-based constructions reviewed here provide important background for the division algebra and Clifford algebra three-generation mechanisms compared in Section~\ref{subsec:division-clifford-comparison}.

\subsection{Topological representations and algebraic spinors}

A parallel line of research represents matter not only algebraically but also topologically. In the model proposed by Bilson--Thompson, first-generation quarks and leptons are represented by framed three-strand braids, with twisting and braiding encoding electric charge, colour-related information, chirality, and aspects of electroweak interactions \cite{bilson2005topological}. Subsequent work established a concrete correspondence between these topological states and algebraic spinors constructed using division algebras and Clifford algebras. In particular, basis states spanning minimal left ideals of $\Cl{6}$ can be mapped to framed-braid states, while the complex and quaternion factors give representations of the corresponding circular braid groups \cite{gresnigt2018braids,gresnigt2019braided}. The correspondence was further developed by identifying the moves that exchange braiding and twisting and by assigning canonical braid representatives to the relevant states \cite{gresnigt2019combing}.

This relation was later extended to include chiral weak structure by combining the $\Cl{6}$ electrocolour ideals with minimal one-sided ideals of $\Cl{4}$. Maps from the respective Witt bases to the circular braid group $B_3^c$ and its subgroup $B_3$ associate fermion states and charged weak generators with braids, while weak transitions are represented by braid composition \cite{gresnigt2020topological}. Conversely, several restrictions originally imposed on the ribbon configurations can be understood as consequences of the nilpotent ladder operators, exterior-algebra structure, and primitive idempotents used to construct the corresponding spinors \cite{gresnigt2021topological}. More recently, Chester, Arsiwalla, and Kauffman related the braid-based states directly to Lie-algebra representation theory, associating ribbon twists with the weight structure of $\SU(3)_C\times\U(1)_{\rm em}$ and the braid data with chiral Lorentz-spinor states \cite{chester2025preons}. Although these topological constructions do not by themselves provide a complete SM gauge structure or a mechanism for three generations, they show that the same particle-state information can admit closely related algebraic and topological descriptions.

\subsection{Octonion dimensional reduction and three leptonic generations}
\label{sec:manogue-dray-dimensional-reduction}

An important early appearance of a three-generation structure in physics based on octonions was the dimensional-reduction mechanism developed by Manogue and Dray \cite{manogue1999dimensional}. They formulate a ten-dimensional massless Dirac equation using two-component spinors over $\bb{O}$ and $2\times 2$ Hermitian matrices over $\bb{O}$, with the corresponding Lorentz transformations described in terms of $SL(2,\bb{O})$. The key step is to choose a preferred imaginary octonion unit $\ell$. This defines a projection onto a preferred complex subalgebra $\bb{C}\subset\bb{O}$ and selects an $SL(2,\bb{C})$ subgroup of $SL(2,\bb{O})$, thereby reducing the Lorentz structure from ten dimensions to four without compactification.

The same choice of $\ell$ selects three distinguished quaternion subalgebras of $\bb{O}$ containing the preferred complex subalgebra. Manogue and Dray interpreted the corresponding sectors as three generations of leptons. The conceptual importance of this construction for the present review is that a preferred complex direction in $\bb{O}$ naturally organises three quaternion subalgebras. This observation subsequently motivated the question of whether three octonion subalgebras of the sedenions could play an analogous role for three complete fermion generations. The relation between this historical precursor and the later sedenion-motivated construction is examined in Section~\ref{subsec:dimensional-reduction-comparison}.

\subsection{Other approaches based on Clifford algebras and exceptional structures}

A broad literature uses Clifford and geometric algebras to organise SM fermions and internal symmetries. Chisholm and Farwell studied gauge transformations of spinors within Clifford algebras \cite{chisholm1996properties,chisholm1999gauge}, while Trayling and Baylis developed a formulation of the SM gauge group using the real geometric algebra $C\ell(7,0)$ of seven-dimensional Euclidean space \cite{trayling1999geometric,trayling2001geometric,trayling2004cl}. Other representative examples include the unified SM construction of Gording and Schmidt--May, Lasenby's analysis of $\SU(3)$ and octonions using geometric algebra, Pav\v{s}i\v{c}'s treatment of quantised fields using Clifford algebras, and more recent models based on larger geometric algebras or mutually commuting Clifford actions \cite{gording2020unified,lasenby2024some,pavsic2017quantized,hamilton2023unification,barrett2024commuting}. Wilson has emphasised the related problem of selecting physically relevant subgroups from the many possibilities contained in a sufficiently large Clifford algebra \cite{wilson2020group,wilson2021problem,wilson2022remarks}.

These works illustrate both the power and the flexibility of Clifford algebras as a language for internal symmetries. The same flexibility creates a selection problem, since a large Clifford algebra generally contains many possible subgroups, ideals, and automorphisms. The framework reviewed here instead obtains $\Cl{8}$ as the associative endomorphism algebra generated by complexified sedenion left-multiplication maps, and constructs the family action from the intrinsic $S_3$ factor in the sedenion automorphism structure. This does not determine every subsequent algebraic choice, but it provides a reason for focusing on $\Cl{8}$ and on an $S_3$ action with a family interpretation.

A separate but related literature uses exceptional Jordan algebras and exceptional Lie groups. The exceptional Jordan algebra $J_3(\bb{O})$, formed from $3\times3$ Hermitian matrices over $\bb{O}$, combines a natural threefold structure with exceptional symmetries. Representative developments include the exceptional-geometry constructions of Dubois-Violette, Todorov, and collaborators; the Jordan-geometric and triality-based proposals of Boyle and Farnsworth; and the $E_6$ and $E_8$ constructions of Manogue, Dray, and Wilson \cite{dubois2016exceptional1,dubois2019exceptional2,todorov2018octonions,todorov2018deducing,boyle2020standard,boyle2020standard2,manogue2022octions,dray2024new}. Perelman has considered models combining Jordan and Clifford algebras, Singh has explored possible connections with fermion masses, and Chester et al.\ have investigated relations between the Dixon and Jordan frameworks \cite{perelman2021jordan,singh2021characteristic,chester2023dixon}. More recently, Baez and Schwahn characterised the SM gauge group through the stabilisers of nested complex Jordan subalgebras inside $J_3(\bb{O})$ \cite{baez2026standard}. These approaches provide important alternative settings for algebraic threefold structures; their proposed generation mechanisms and their relation to the construction reviewed here are discussed in Section~\ref{subsec:jordan-comparison}.

\subsection{$\Cl{8}$, triality, and three-generation mechanisms}

The exceptional outer-automorphism structure of $\Spin(8)$ has long suggested a possible relation between triality and the existence of three fermion generations. One has
\begin{eqnarray}
\operatorname{Out}\bigl(\Spin(8)\bigr)&\simeq&S_3,
\end{eqnarray}
with the order-three element permuting the vector representation and the two inequivalent chiral-spinor representations,
\begin{eqnarray}
\mathbf{8}_v,\qquad \mathbf{8}_s,\qquad \mathbf{8}_c.
\end{eqnarray}
Silagadze was among the first to propose that an $SO(8)$ colour structure and its triality might provide an origin for three generations \cite{silagadze1994so}. More recent triality-based constructions include those of Furey and Hughes, and of Quinta \cite{furey2025three,quinta2025spacetime}. Triality also arises naturally in the exceptional Jordan algebra $J_3(\bb{O})$, where permutations of three distinguished diagonal idempotents induce the outer triality action on the three off-diagonal octonion sectors \cite{todorov2018octonions,boyle2020standard2}.

The $S_3$ family action considered in the present framework should be distinguished from this standard triality action. Standard $\Spin(8)$ triality permutes the inequivalent representations $\mathbf{8}_v$, $\mathbf{8}_s$, and $\mathbf{8}_c$, whereas the family action reviewed here is constructed from the intrinsic $S_3$ structure associated with the sedenions and is realised as an automorphism of the full algebra $\Cl{8}$. It relates three gauge-equivalent fermion sectors rather than identifying the generations directly with the three triality representations. The detailed algebraic reason that these actions do not coincide in the present formulation, together with a comparison of representative triality-based models, is given in Section~\ref{subsec:triality-comparison}. Whether the two $S_3$ structures admit a deeper common interpretation remains an open question considered further in Section~\ref{sec:conclusions-outlook}.

\subsection{From octonion subalgebras to an $S_3$-generated family structure}
\label{sec:historical-development}

The framework reviewed here grew out of the suggestion that three octonion subalgebras of $\bb{S}$ might play a role analogous to the three quaternion sectors identified by Manogue and Dray in $\bb{O}$ \cite{manogue1999dimensional,Gillard2019,gillard2019c,gresnigt2019sedenions}. In these early models, the generation sectors were constructed from three overlapping, generation-adapted $\Cl{6}$ structures inside the sedenions. Although this identified a natural threefold structure, it did not yet provide three linearly independent fermion sectors acted upon by a single common gauge algebra.

This limitation motivated a shift from identifying generations directly with octonion subalgebras to constructing them through the intrinsic $S_3$ structure associated with the sedenions and its realisation within $\Cl{8}$. The later refined $\Cl{8}$ construction is therefore not merely the earlier model rewritten in a different Clifford basis: the direct subalgebra interpretation is replaced by an $S_3$-generated orbit of fermion subspaces within a single associative algebra. Successive developments established an invariant colour symmetry, three linearly independent fermion sectors with a common electromagnetic symmetry, and finally a $\Cl{10}$ extension containing the full SM gauge group as a single common gauge sector \cite{Gresnigt2023,gresnigt2023toward,gresnigt2023sedenion,gourlay2024algebraic,Gresnigt2026PLB}. The following sections reconstruct this development and present the resulting $\Cl{8}$ and $\Cl{10}$ formulations in detail.

\section{Algebraic foundations: from the octonion $\Cl{6}$ construction to sedenion $\Cl{8}$}
\label{sec:algebraic-foundations}

\subsection{The real normed division algebras}

The real normed division algebras occupy a distinguished position among finite-dimensional algebras. Hurwitz's theorem states that there are, up to isomorphism, only four finite-dimensional normed division algebras over $\bb{R}$:
\begin{eqnarray}
\bb{R},\qquad \bb{C},\qquad \bb{H},\qquad \bb{O}
\end{eqnarray}
of real dimensions $1$, $2$, $4$, and $8$, respectively \cite{Baez2002Octonions,Dixon1994}. Each carries a canonical conjugation $x\mapsto\overline{x}$ and a positive-definite quadratic norm
\begin{eqnarray}
N(x)&:=&x\overline{x}=\overline{x}x
\end{eqnarray}
satisfying the composition property
\begin{eqnarray}
N(xy)&=&N(x)N(y).
\end{eqnarray}
For $x\neq0$, this gives the inverse $x^{-1}=\overline{x}/N(x)$. The successive algebras exhibit a corresponding weakening of algebraic structure: The real numbers $\bb{R}$ are self-conjugate, commutative, and associative; the complex numbers $\bb{C}$ remain commutative and associative but acquire a nontrivial involution; the quaternions $\bb{H}$ remain associative but become noncommutative; and the octonions $\bb{O}$ are noncommutative and nonassociative, although still alternative.

The quaternion algebra may be written as
\begin{eqnarray}
\bb{H}&=&\operatorname{span}_{\bb{R}}\{1,\mathbf{i},\mathbf{j},\mathbf{k}\}
\end{eqnarray}
with multiplication determined by
\begin{eqnarray}
\mathbf{i}^2=\mathbf{j}^2=\mathbf{k}^2=\mathbf{i}\mathbf{j}\mathbf{k}&=&-1
\end{eqnarray}
or equivalently
\begin{eqnarray}
\mathbf{i}\mathbf{j}=-\mathbf{j}\mathbf{i}=\mathbf{k},\qquad \mathbf{j}\mathbf{k}=-\mathbf{k}\mathbf{j}=\mathbf{i},\qquad \mathbf{k}\mathbf{i}=-\mathbf{i}\mathbf{k}=\mathbf{j}.
\end{eqnarray}
For
\begin{eqnarray}
q&=&q_0+q_1\mathbf{i}+q_2\mathbf{j}+q_3\mathbf{k}
\end{eqnarray}
the canonical conjugation and norm are
\begin{eqnarray}
\overline{q}&=&q_0-q_1\mathbf{i}-q_2\mathbf{j}-q_3\mathbf{k},\qquad
N(q)=q\overline{q}=q_0^2+q_1^2+q_2^2+q_3^2.
\end{eqnarray}
Automorphisms of $\bb{H}$ fix the real line and act by rotations on the three-dimensional imaginary subspace. Consequently,
\begin{eqnarray}
\Aut(\bb{H})&\simeq&\SO(3).
\end{eqnarray}
More explicitly, every automorphism is realised by conjugation with a unit quaternion,
\begin{eqnarray}
q&\longmapsto&uqu^{-1},\qquad N(u)=1,
\end{eqnarray}
and the unit quaternions form $\SU(2)$, with the elements $u$ and $-u$ inducing the same automorphism. Thus $\SU(2)$ provides the familiar double cover of $\SO(3)$.

The octonions form the largest of the four normed division algebras. We write
\begin{eqnarray}
\bb{O}&=&\operatorname{span}_{\bb{R}}\{1,o_1,\ldots,o_7\}.
\end{eqnarray}
To fix conventions for the remainder of the paper, their multiplication is written compactly as
\begin{eqnarray}
o_i o_j&=&-\delta_{ij}+\varphi_{ijk}o_k,\qquad i,j=1,\ldots,7,
\label{eq:octonion-multiplication}
\end{eqnarray}
where $\varphi_{ijk}$ is totally antisymmetric and $\varphi_{ijk}=+1$ for the seven oriented triples
\begin{eqnarray}
(1,2,3),\quad (1,4,5),\quad (1,7,6),\quad (2,4,6),\quad (2,5,7),\quad (3,4,7),\quad (3,6,5).
\end{eqnarray}
This convention agrees with the octonion subalgebra of the sedenions used below, under the identification $o_i\leftrightarrow s_i$ for $i=1,\ldots,7$. 

\begin{figure}[htbp]
\centering
\includegraphics[width=0.35\textwidth]{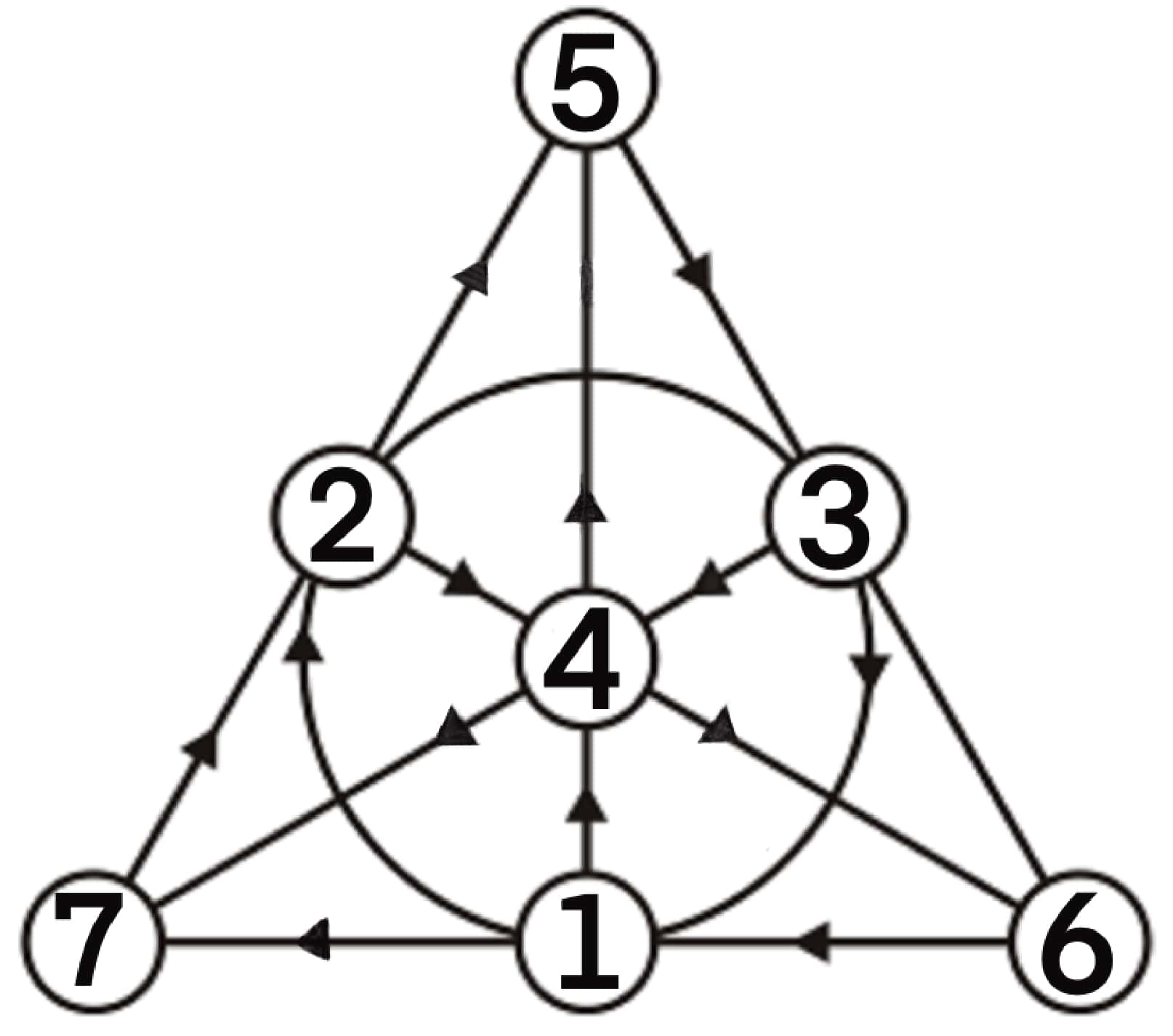}
\caption{The Fano-plane representation of the octonion multiplication convention used in this review. Each oriented line corresponds to one associative quaternionic triple of imaginary octonion units.}
\label{fig:fano-plane}
\end{figure}

The seven oriented triples may be represented geometrically by the Fano plane shown in Figure~\ref{fig:fano-plane}. Each oriented line determines the multiplication of the three corresponding imaginary octonion units; reversing the orientation changes the sign of the product.

For
\begin{eqnarray}
x&=&x_0+\sum_{i=1}^{7}x_i o_i
\end{eqnarray}
one has
\begin{eqnarray}
\overline{x}&=&x_0-\sum_{i=1}^{7}x_i o_i,\qquad
N(x)=x\overline{x}=x_0^2+\sum_{i=1}^{7}x_i^2.
\end{eqnarray}

Although $\bb{O}$ is nonassociative, it is alternative. Writing the associator as
\begin{eqnarray}
(x,y,z)&:=&(xy)z-x(yz),
\end{eqnarray}
alternativity means
\begin{eqnarray}
(x,x,y)&=&0,\qquad (y,x,x)=0.
\end{eqnarray}
In particular, every subalgebra generated by two octonions is associative. 

The automorphism group of the octonions is the compact exceptional Lie group
\begin{eqnarray}
\Aut(\bb{O})&=&G_2.
\end{eqnarray}
It preserves both the octonion multiplication and the norm. The stabiliser in $G_2$ of a chosen unit imaginary octonion is isomorphic to $\SU(3)$,
\begin{eqnarray}
\operatorname{Stab}_{G_2}(u)&\simeq&\SU(3),
\qquad u\in\operatorname{Im}\bb{O},\qquad N(u)=1,
\end{eqnarray}
a fact that underlies the recurring connection between octonions and colour symmetry reviewed in Section~\ref{sec:historical-remarks}.

The division algebras also contain a natural hierarchy of lower-dimensional subalgebras. For the standard quaternion basis, the three imaginary units determine three distinguished complex subalgebras,
\begin{eqnarray}
\operatorname{span}_{\bb{R}}\{1,\mathbf{i}\},\qquad
\operatorname{span}_{\bb{R}}\{1,\mathbf{j}\},\qquad
\operatorname{span}_{\bb{R}}\{1,\mathbf{k}\},
\end{eqnarray}
all sharing the same real line. Similarly, the seven oriented octonion triples introduced above determine seven distinguished basis-generated quaternionic subalgebras,
\begin{eqnarray}
\bb{H}_{abc}
&:=&
\operatorname{span}_{\bb{R}}\{1,o_a,o_b,o_c\}
\subset
\bb{O},
\end{eqnarray}
one for each line of the Fano plane.

These quaternionic subalgebras overlap along distinguished complex subalgebras. Each standard imaginary octonion unit belongs to exactly three of the seven quaternionic copies. For example, $o_1$ occurs in the triples $(1,2,3)$, $(1,4,5)$, and $(1,7,6)$, so that
\begin{eqnarray}
\bb{H}_{123},\qquad
\bb{H}_{145},\qquad
\bb{H}_{176},
\end{eqnarray}
all contain the common complex subalgebra
\begin{eqnarray}
\operatorname{span}_{\bb{R}}\{1,o_1\}
&\simeq&
\bb{C}.
\end{eqnarray}
Thus a chosen imaginary octonion direction naturally selects three quaternionic subalgebras containing the same copy of $\bb{C}$. This observation underlies the three-sector construction of Manogue and Dray reviewed in Section~\ref{sec:manogue-dray-dimensional-reduction}, and the same nested complex--quaternionic--octonion pattern reappears in enlarged form inside the sedenions.

It is worth emphasising that algebras such as $\bb{C}\otimes\bb{H}$, $\bb{C}\otimes\bb{O}$, and $\bb{C}\otimes\bb{H}\otimes\bb{O}$ are not division algebras and contain nontrivial idempotents. For example, if $u^2=-1$ is an imaginary unit and $\ii$ denotes the commuting complex unit, then
\begin{eqnarray}
\rho_\pm
&:=&
\frac{1}{2}\left(1\pm\ii u\right),
\end{eqnarray}
satisfy
\begin{eqnarray}
\rho_\pm^2=\rho_\pm,\qquad
\rho_+\rho_-&=&0.
\end{eqnarray}
In particular, Dixon's algebra
\begin{eqnarray}
\bb{T}
&=&
\bb{C}\otimes\bb{H}\otimes\bb{O},
\end{eqnarray}
is nonassociative and nonalternative \cite{Dixon1994,Dixon2010}. 

\subsubsection{The Cayley--Dickson doubling process}
The four normed division algebras are not unrelated isolated structures. They may be generated successively by the Cayley--Dickson doubling process \cite{Baez2002Octonions,Dixon1994},
\begin{eqnarray}
\bb{R}\longrightarrow\bb{C}\longrightarrow\bb{H}\longrightarrow\bb{O},
\end{eqnarray}
Let $A_n$ be a real algebra equipped with a conjugation $a\mapsto\overline{a}$. The next algebra is the real vector space
\begin{eqnarray}
A_{n+1}&=&A_n\oplus A_n\ell_n,
\end{eqnarray}
whose elements may be written as
\begin{eqnarray}
(a,b)&\equiv&a+b\ell_n,\qquad a,b\in A_n.
\end{eqnarray}
The product and conjugation are defined by
\begin{eqnarray}
(a,b)(c,d)&=&\left(ac-\overline{d}b,\;da+b\overline{c}\right),\\
\overline{(a,b)}&=&(\overline{a},-b).
\end{eqnarray}
In particular,
\begin{eqnarray}
\ell_n^2&=&-1,\qquad \ell_n a=\overline{a}\,\ell_n,
\end{eqnarray}
and the associated quadratic form satisfies
\begin{eqnarray}
N(a,b)&:=&(a,b)\overline{(a,b)}=N(a)+N(b).
\end{eqnarray}

Starting from $A_0=\bb{R}$, successive doublings give
\begin{eqnarray}
A_0=\bb{R}\quad
\longrightarrow \quad
A_1=\bb{C}\quad
\longrightarrow \quad
A_2=\bb{H} \quad
\longrightarrow \quad
A_3=\bb{O} \quad
\longrightarrow \quad
A_4=\bb{S}.
\end{eqnarray}
For $A_0,\ldots,A_3$ the quadratic form is multiplicative, giving precisely the four normed division algebras identified by Hurwitz's theorem. The Cayley--Dickson process itself does not terminate at $\bb{O}$, however. The next doubling produces the sixteen-dimensional sedenion algebra $\bb{S}$, for which the composition and division properties are lost. This first step beyond the octonions is the subject of the next subsection.

\subsection{The sedenions}
\label{sec:sedenions}

The sedenions $\bb{S}$ are obtained by applying one further Cayley--Dickson doubling step to the octonions,
\begin{eqnarray}
\bb{S}
&=&
\bb{O}\oplus\bb{O}s_8,
\end{eqnarray}
and therefore have real dimension sixteen. A general sedenion may be written as
\begin{eqnarray}
w
&=&
w_0s_0+w_1s_1+\cdots+w_{15}s_{15},
\qquad
w_i\in\bb{R},
\end{eqnarray}
where $s_0=1$ and $s_1,\ldots,s_{15}$ are imaginary basis elements satisfying
\begin{eqnarray}
s_i^2&=&-1,
\qquad
s_i s_j=-s_j s_i,
\qquad
i\neq j.
\end{eqnarray}
The first seven imaginary units $s_1,\ldots,s_7$ span the distinguished octonion subalgebra inherited from the Cayley--Dickson construction and are identified with the octonion basis introduced above.

As for the octonions, the multiplication of imaginary basis elements may be written compactly as
\begin{eqnarray}
s_i s_j
&=&
-\delta_{ij}+C_{ijk}s_k,
\qquad
i,j=1,\ldots,15,
\end{eqnarray}
where $C_{ijk}$ are totally antisymmetric structure constants determined by the chosen Cayley--Dickson convention. Their restriction to $i,j,k\in\{1,\ldots,7\}$ reproduces the octonion structure constants $\varphi_{ijk}$ introduced above. The complete multiplication convention used throughout this review is given in Appendix~\ref{app:sedenion-multiplication-table}.

The standard conjugation is
\begin{eqnarray}
\overline{w}
&=&
w_0s_0-w_1s_1-\cdots-w_{15}s_{15},
\end{eqnarray}
and the associated quadratic form is
\begin{eqnarray}
N(w)
&:=&
w\overline{w}
=
\overline{w}w
=
\sum_{i=0}^{15}w_i^2.
\end{eqnarray}
The quadratic form remains positive definite, but it is no longer multiplicative:
\begin{eqnarray}
N(xy)&\neq&N(x)N(y)
\end{eqnarray}
in general. Consequently, $\bb{S}$ is no longer a composition algebra or a division algebra and contains nonzero zero divisors. The sedenions are also nonalternative, although they retain the weaker properties of power-associativity and flexibility. Power-associativity means that the subalgebra generated by any single element is associative, so that powers $x^n$ are unambiguously defined, while flexibility means
\begin{eqnarray}
(xy)x&=&x(yx),
\end{eqnarray}
or equivalently $(x,y,x)=0$ for the associator $(x,y,z)=(xy)z-x(yz)$.

The zero divisors of $\bb{S}$ nevertheless possess considerable structure. Moreno gave an algebraic description of zero divisors in the real Cayley--Dickson algebras $A_n$ for $n\geq4$, while Baez noted that the space of unit-norm zero divisors in $\bb{S}$ is closely related to the compact exceptional group $G_2$ \cite{Moreno1998ZeroDivisors,Baez2002Octonions}. At the same time, the real sedenion algebra contains no nontrivial idempotents \cite{BresarSemrlSpenko2011LocallyComplex}; its zero divisors should therefore not be confused with the primitive idempotents used later to construct Clifford spinors.

Of particular importance for the present framework is the automorphism structure. Brown \cite{Brown1967GeneralizedCayleyDickson} showed that the automorphism group of the standard Cayley--Dickson sedenions is
\begin{eqnarray}
\Aut(\bb{S})&\simeq&G_2\times S_3.
\end{eqnarray}
The $G_2$ factor extends the familiar automorphism symmetry of the octonions, while the additional discrete $S_3$ factor is a genuinely new feature appearing at the sedenion stage. It is this $S_3$ factor that will provide the candidate family symmetry developed below.

Although the Cayley--Dickson construction may be continued indefinitely beyond $\bb{O}$, the sedenions are distinguished as the first step at which a new finite automorphism structure appears. In particular,
\begin{eqnarray}
\Aut(\bb{S})
&\simeq&
G_2\times S_3.
\end{eqnarray}
The continuous part therefore retains the exceptional $G_2$ symmetry already present for the octonions, while the additional $S_3$ factor is new at the sedenion stage. For the subsequent standard Cayley--Dickson algebras, the continuous derivation algebra remains $\mathfrak{g}_2$, while further doublings introduce additional discrete $S_3$ factors \cite{schafer1954algebras,Brown1967GeneralizedCayleyDickson}. Thus $\bb{S}$ is the minimal Cayley--Dickson extension of $\bb{O}$ in which the new discrete structure relevant to a three-family interpretation is already present.

The standard Cayley--Dickson basis also exhibits a rich hierarchy of lower-dimensional subalgebras. Besides the distinguished octonion subalgebra
\begin{eqnarray}
\bb{O}_1
&:=&
\operatorname{span}_{\bb{R}}\{1,s_1,\ldots,s_7\},
\end{eqnarray}
the sedenion basis contains seven further basis-generated copies of $\bb{O}$ \cite{Cawagas2004}. These octonion subalgebras are not disjoint: distinct copies intersect along distinguished quaternionic subalgebras, while suitable triple intersections may reduce further to a common complex subalgebra. For example,
\begin{eqnarray}
\bb{O}_3\cap\bb{O}_5\cap\bb{O}_7
&=&
\operatorname{span}_{\bb{R}}\{1,s_4,s_8,s_{12}\}
\simeq
\bb{H},
\end{eqnarray}
whereas
\begin{eqnarray}
\bb{O}_1\cap\bb{O}_2\cap\bb{O}_3
&=&
\operatorname{span}_{\bb{R}}\{1,s_1\}
\simeq
\bb{C}.
\end{eqnarray}
This overlapping octonion structure provided the original motivation for associating three fermion generations with three octonion subalgebras of $\bb{S}$. As discussed in Section~\ref{sec:historical-remarks}, the later formulation no longer identifies generations directly with these subalgebras, but instead uses the intrinsic $S_3$ automorphism structure of $\bb{S}$.

\subsubsection{The intrinsic $S_3$ automorphisms of the sedenions}

It is useful to recall the explicit form of the sedenion automorphisms. Writing a sedenion as
\begin{eqnarray}
A+Bs_8,\qquad A,B\in\bb{O},
\end{eqnarray}
one obtains automorphisms of the form
\begin{eqnarray}
\varphi:A+Bs_8 &\mapsto& \varphi(A)+\varphi(B)s_8,
\end{eqnarray}
where $\varphi\in G_2=\operatorname{Aut}(\bb{O})$. In addition, there are two discrete transformations
\begin{eqnarray}
\epsilon:A+Bs_8 &\mapsto& A-Bs_8,
\label{eq:sedenion-order-two-action}
\end{eqnarray}
and
\begin{eqnarray}
\psi:A+Bs_8
&\mapsto&
\frac{1}{4}\left[A+3A^\ast+\sqrt{3}(B-B^\ast)\right]
+
\frac{1}{4}\left[B+3B^\ast-\sqrt{3}(A-A^\ast)\right]s_8,
\label{eq:sedenion-order-three-action}
\end{eqnarray}
where $A^\ast=\overline{A}$ denotes the standard octonion conjugation on the octonion factor in the Cayley--Dickson decomposition $\bb{S}=\bb{O}\oplus\bb{O}\ell$. These maps satisfy
\begin{eqnarray}
\epsilon^2 &=& \operatorname{Id}, \qquad
\psi^3 = \operatorname{Id}, \qquad
\epsilon\psi = \psi^2\epsilon,
\end{eqnarray}
and therefore generate $S_3$. Since octonion automorphisms $\varphi$ commute with the standard octonion conjugation, the componentwise $G_2$ action on $\bb{S}=\bb{O}\oplus\bb{O}s_8$ commutes with both discrete automorphisms $\epsilon$ and $\psi$.

It is this discrete automorphism structure that supplies the candidate family symmetry in the construction reviewed below. The order-three element is particularly important. In a suitable basis its action mixes the original octonion directions with the new sedenion directions. Schematically,
\begin{eqnarray}
\psi(s_i) &\sim& -\frac{1}{2}s_i-\frac{\sqrt{3}}{2}s_{i+8}, \quad
\psi(s_{i+8}) \sim -\frac{1}{2}s_{i+8}+\frac{\sqrt{3}}{2}s_i,
\qquad i=1,\ldots,7.
\end{eqnarray}
Thus $\psi$ is not an automorphism of the original octonion subalgebra alone. It is a symmetry specific to the sedenions. This is crucial: if three generations are to arise from an operation unavailable inside $\bb{O}$, then the step from $\bb{O}$ to $\bb{S}$ has a clear structural purpose.

In the later Clifford algebra formulation, the corresponding order-three transformation is denoted by $\psi_3$, emphasising its role as the cyclic generator of the family action. The order-two generator is denoted by $\epsilon$. The next step is to realise an $S_3$ action motivated by these sedenion automorphisms within the associated $\Cl{8}$ framework. Once this has been done, one can study its action on the primitive idempotents, minimal ideals, semi-spinor spaces, and fermion subspaces of $\Cl{8}$.

The significance of this discrete symmetry for the present framework is immediate. The SM contains three fermion generations with identical gauge quantum numbers, suggesting that a family symmetry should relate three algebraically distinct fermion sectors without simultaneously triplicating the gauge sector. The intrinsic $S_3$ of $\bb{S}$ provides a natural candidate for such a symmetry: its order-three element can generate a three-element family orbit, while the common $G_2$ structure retains the octonion symmetry associated with colour.

The central idea of the construction reviewed below is therefore not to identify the three generations directly with three octonion subalgebras of $\bb{S}$, but to realise the intrinsic sedenion $S_3$ as an action on fermionic spinor spaces inside an associated Clifford algebra, in such a way that it results in three linearly independent fermion families. In this way the family symmetry, fermion states, and gauge generators can ultimately be represented within a single associative algebraic structure, while the selected gauge sector remains untriplicated.

\subsection{From multiplication maps to Clifford algebras}
\label{sec:multiplication-algebras}

The division and Cayley--Dickson algebras introduced above are not themselves Clifford algebras. The relevant Clifford algebras arise instead as associative algebras of endomorphisms generated by their left- and right-multiplication maps. This distinction is particularly important for the nonassociative algebras $\bb{O}$ and $\bb{S}$. Although multiplication in these algebras is nonassociative, their multiplication maps are linear endomorphisms, and compositions of these maps are associative under composition. Moreover, a composition of multiplication maps need not itself be multiplication by a single algebra element, allowing the resulting endomorphism algebra to be much larger than the original algebra \cite{Dixon1994,Furey2026Nested}

Let $\bb{A}$ be an algebra over a field $\bb{F}$. For each $a\in\bb{A}$, define the left and right multiplication maps
\begin{eqnarray}
L_a[x]&:=&ax,
\qquad
R_a[x]:=xa,
\qquad
x\in\bb{A}.
\end{eqnarray}
The left multiplication algebra $\bb{A}_L$ is the subalgebra of $\operatorname{End}_{\bb{F}}(\bb{A})$ generated by the maps $L_a$, and similarly $\bb{A}_R$ is generated by the maps $R_a$. We denote by
\begin{eqnarray}
\operatorname{Mult}(\bb{A})
&\subseteq&
\operatorname{End}_{\bb{F}}(\bb{A})
\end{eqnarray}
the full multiplication algebra generated by both left and right multiplication maps.

The product in each of these multiplication algebras is composition of linear maps and is therefore associative, irrespective of whether $\bb{A}$ itself is associative. For example,
\begin{eqnarray}
L_aL_b[x]&=&a(bx),
\qquad
R_aR_b[x]=(xb)a.
\end{eqnarray}
When $\bb{A}$ is nonassociative, one generally has
\begin{eqnarray}
L_aL_b&\neq&L_{ab},
\qquad
R_aR_b\neq R_{ba}.
\end{eqnarray}
Consequently, repeated multiplication can generate an associative action algebra of substantially larger dimension than the original nonassociative algebra.

For the real normed division algebras these multiplication algebras are closely related to low-dimensional real Clifford algebras. We use the convention that $C\ell(p,q)$ has $p$ generators squaring to $+1$ and $q$ generators squaring to $-1$. For the complex numbers,
\begin{eqnarray}
\bb{C}_L
\simeq
\bb{C}_R
\simeq
\bb{C}
\simeq
C\ell(0,1),
\end{eqnarray}
where $\bb{C}$ is regarded here as a real algebra. For the quaternions the left and right multiplication algebras are distinct copies of
\begin{eqnarray}
\bb{H}_L
\simeq
\bb{H}_R
\simeq
\bb{H}
\simeq
C\ell(0,2).
\end{eqnarray}
Because $\bb{H}$ is associative, the left and right actions commute, and together they generate the full real endomorphism algebra,
\begin{eqnarray}
\operatorname{Mult}(\bb{H})
&\simeq&
\bb{H}_L\otimes_{\bb{R}}\bb{H}_R
\simeq
\operatorname{End}_{\bb{R}}(\bb{H})
\simeq
\Mat(4,\bb{R})
\simeq
C\ell(3,1).
\end{eqnarray}

The octonion case is qualitatively different. Because $\bb{O}$ is nonassociative, compositions of left multiplication maps already generate the full endomorphism algebra:
\begin{eqnarray}
\bb{O}_L
\simeq
\bb{O}_R
\simeq
\operatorname{Mult}(\bb{O})
&\simeq&
\operatorname{End}_{\bb{R}}(\bb{O})
\simeq
\Mat(8,\bb{R})
\simeq
C\ell(0,6).
\end{eqnarray}
Thus the eight-dimensional nonassociative algebra $\bb{O}$ gives rise, through composed multiplication maps, to the $64$-dimensional associative algebra $C\ell(0,6)$.

The sedenions provide the next example relevant to this review. As shown explicitly in Section~\ref{sec:CS-to-Cl8}, eight distinguished sedenion left multiplication maps satisfy the defining relations of $C\ell(0,8)$ and generate
\begin{eqnarray}
\bb{S}_L
&\simeq&
\operatorname{End}_{\bb{R}}(\bb{S})
\simeq
\Mat(16,\bb{R})
\simeq
C\ell(0,8).
\end{eqnarray}
The fact that $\bb{S}$ itself is nonalternative therefore presents no obstacle to obtaining an associative Clifford algebra of linear actions.

Upon complexification, the signature of a Clifford algebra becomes irrelevant:
\begin{eqnarray}
\bb{C}\otimes_{\bb{R}}C\ell(p,q)
&\simeq&
\Cl{p+q}.
\end{eqnarray}
The cases needed repeatedly below are consequently
\begin{eqnarray}
(\bb{C}\otimes\bb{H})_L
&\simeq&
\Cl{2},\\
(\bb{C}\otimes\bb{O})_L
&\simeq&
\Cl{6},\\
(\bb{C}\otimes\bb{S})_L
&\simeq&
\Cl{8}.
\end{eqnarray}
For the complexified quaternions, including both left and right multiplication enlarges the action algebra to
\begin{eqnarray}
\operatorname{Mult}_{\bb{C}}(\bb{C}\otimes\bb{H})
&\simeq&
\Cl{4}.
\end{eqnarray}

These structures also explain the appearance of $\Cl{10}$ in constructions based on the combined quaternionic and octonion sectors. The complex vector space
\begin{eqnarray}
\bb{C}\otimes\bb{H}\otimes\bb{O}
\end{eqnarray}
has complex dimension $32$, and its full multiplication algebra is
\begin{eqnarray}
\operatorname{Mult}_{\bb{C}}
\left(
\bb{C}\otimes\bb{H}\otimes\bb{O}
\right)
&\simeq&
\operatorname{End}_{\bb{C}}
\left(
\bb{C}\otimes\bb{H}\otimes\bb{O}
\right)
\simeq
\Mat(32,\bb{C})
\simeq
\Cl{10}.
\end{eqnarray}
Equivalently, the six Clifford directions generated by the complex octonion multiplication algebra combine with the four directions generated by the left and right complex quaternionic actions,
\begin{eqnarray}
\Cl{6}\,\widehat{\otimes}\,\Cl{4}
&\simeq&
\Cl{10}.
\end{eqnarray}
This provides the multiplication-algebra perspective underlying earlier constructions based on $\bb{C}\otimes\bb{H}\otimes\bb{O}$ and the five-pair ladder-operator formulation of $\Cl{10}$ \cite{Furey2018a}, as well as the subsequent ideal-based $\Cl{10}$ construction in which the separate $\Cl{6}$ and $\Cl{4}$ structures of the physical states are retained \cite{Gresnigt2020StandardModel}.

The general principle is therefore simple: the underlying division or Cayley--Dickson algebra supplies distinguished algebraic structure, while its multiplication maps generate a larger associative algebra in which Clifford spinors and internal symmetry generators can be represented. In the remainder of this section we use this principle first in the familiar one-generation case
\begin{eqnarray}
\bb{C}\otimes\bb{O}
&\longrightarrow&
\Cl{6},
\end{eqnarray}
and subsequently in the sedenion case
\begin{eqnarray}
\bb{C}\otimes\bb{S}
&\longrightarrow&
\Cl{8}.
\end{eqnarray}

\subsection{A $\Cl{6}$ construction for one fermion generation based on octonions}
\label{sec:Cl6-one-generation-template}

Having established the relation between multiplication algebras and Clifford algebras, we now review a $\Cl{6}$ construction based on the complex octonions that describes the electrocolour quantum numbers of one fermion generation. The formulation used here follows \cite{furey2016standard}; its antecedents in octonion colour models and fermionic-oscillator constructions were discussed in Section~\ref{sec:historical-remarks}.

As reviewed in Section~\ref{sec:multiplication-algebras}, the complex octonion left multiplication algebra is
\begin{eqnarray}
(\bb{C}\otimes\bb{O})_L
\simeq
\Mat(8,\bb{C})
\simeq
\Cl{6}.
\end{eqnarray}
Six independent octonion left actions may therefore be chosen as Clifford generators
\begin{eqnarray}
\gamma_i&:=&L_{o_i},
\qquad
i=1,\ldots,6,
\end{eqnarray}
satisfying
\begin{eqnarray}
\{\gamma_i,\gamma_j\}
&=&
-2\delta_{ij}\mathbf{1}_8.
\end{eqnarray}
The remaining imaginary octonion left action is contained in the algebra generated by these six maps and is represented, up to convention-dependent sign, by their sixfold Clifford product.

A convenient Witt basis is
\begin{eqnarray}
\alpha_1
&=&
\frac{1}{2}\left(-\gamma_5+\ii\gamma_4\right),
\qquad
\alpha_1^\dagger
=
\frac{1}{2}\left(\gamma_5+\ii\gamma_4\right),\\
\alpha_2
&=&
\frac{1}{2}\left(-\gamma_3+\ii\gamma_1\right),
\qquad
\alpha_2^\dagger
=
\frac{1}{2}\left(\gamma_3+\ii\gamma_1\right),\\
\alpha_3
&=&
\frac{1}{2}\left(-\gamma_6+\ii\gamma_2\right),
\qquad
\alpha_3^\dagger
=
\frac{1}{2}\left(\gamma_6+\ii\gamma_2\right).
\end{eqnarray}
These satisfy the fermionic oscillator relations
\begin{eqnarray}
\{\alpha_i,\alpha_j\}
=
\{\alpha_i^\dagger,\alpha_j^\dagger\}
&=&
0,
\qquad
\{\alpha_i,\alpha_j^\dagger\}
=
\delta_{ij},
\end{eqnarray}
and span complementary maximal totally isotropic subspaces
\begin{eqnarray}
W
:=
\operatorname{span}_{\bb{C}}\{\alpha_1,\alpha_2,\alpha_3\},\quad
W^\dagger
:=
\operatorname{span}_{\bb{C}}\{\alpha_1^\dagger,\alpha_2^\dagger,\alpha_3^\dagger\}.
\end{eqnarray}

Defining
\begin{eqnarray}
\omega
&:=&
\alpha_1\alpha_2\alpha_3,
\qquad
f:=\omega\omega^\dagger,
\qquad
f^c:=\omega^\dagger\omega,
\end{eqnarray}
gives two conjugate primitive idempotents.\footnote{An idempotent is said to be primitive if it is nonzero and cannot be decomposed as a sum of two nonzero orthogonal idempotents.} Since
\begin{eqnarray}
\alpha_i f&=&0,
\qquad
i=1,2,3,
\end{eqnarray}
the idempotent $f$ plays the role of a Clifford vacuum. The operators $\alpha_i$ annihilate this vacuum, while the nilpotent and mutually anticommuting operators $\alpha_i^\dagger$ act as fermionic creation operators,
\begin{eqnarray}
(\alpha_i^\dagger)^2&=&0,
\qquad
\alpha_i^\dagger\alpha_j^\dagger=-\alpha_j^\dagger\alpha_i^\dagger.
\end{eqnarray}
They therefore generate the exterior algebra $\Lambda^\bullet W^\dagger$. The corresponding conjugate minimal left ideals are
\begin{eqnarray}
\mathcal I
&:=&
\Cl{6}f
\simeq
\Lambda^\bullet W^\dagger f,
\qquad
\mathcal I^c
:=
\Cl{6}f^c
\simeq
\Lambda^\bullet W f^c,
\end{eqnarray}
each of complex dimension eight. Complex conjugation $\ii\mapsto-\ii$ exchanges the two Witt systems, the primitive idempotents, and hence the two ideals. Although $\Cl{6}$ contains additional minimal left ideals, these two conjugate ideals are selected in this construction because together they furnish the sixteen particle and antiparticle states of one SM generation.

The Witt decomposition
\begin{eqnarray}
V_{\bb{C}}
&=&
W\oplus W^\dagger,
\end{eqnarray}
selects a complex structure on the six-dimensional Clifford generating space. The full bivector algebra is $\spin(6)\simeq\mathfrak{su}(4)$, but a generic bivector need not preserve the two maximal totally isotropic subspaces separately. Let $G\in\Cl{6}$ denote a Hermitian generator corresponding, up to the conventional factor of $\ii$, to an element of $\spin(6)$. Requiring
\begin{eqnarray}
[G,W]
&\subseteq&
W,
\qquad
[G,W^\dagger]
\subseteq
W^\dagger,
\qquad
G^\dagger=G,
\end{eqnarray}
restricts $\spin(6)$ to the stabiliser of the chosen Witt decomposition, namely
\begin{eqnarray}
\mathfrak{u}(3)
&\subset&
\spin(6).
\end{eqnarray}
More generally, for a complex Clifford algebra $\Cl{2n}$ equipped with a chosen $n$-pair Witt decomposition, the unitary transformations preserving the two maximal totally isotropic subspaces separately form $\mathfrak{u}(n)$.

In the present construction $\mathfrak{u}(3)$ is identified as
\begin{eqnarray}
\mathfrak{u}(3)
&=&
\mathfrak{su}(3)_C
\oplus
\mathfrak{u}(1)_{\rm em}.
\end{eqnarray}

Writing
\begin{eqnarray}
\boldsymbol{\alpha}
&:=&
(\alpha_1,\alpha_2,\alpha_3)^T,
\end{eqnarray}
the colour and electromagnetic generators may be written as
\begin{eqnarray}
\Lambda_a
&:=&
-\boldsymbol{\alpha}^\dagger\lambda_a^T\boldsymbol{\alpha},
\qquad
a=1,\ldots,8,\\
Q
&:=&
\frac{1}{3}
\sum_{i=1}^{3}
\alpha_i^\dagger\alpha_i,
\end{eqnarray}
where $\lambda_a$ are the Gell--Mann matrices. The corresponding decomposition of the two ideals is
\begin{eqnarray}
\mathcal I
&\simeq&
\mathbf{1}_{0}
\oplus
\overline{\mathbf{3}}_{\frac{1}{3}}
\oplus
\mathbf{3}_{\frac{2}{3}}
\oplus
\mathbf{1}_{1},\\
\mathcal I^c
&\simeq&
\mathbf{1}_{0}
\oplus
\mathbf{3}_{-\frac{1}{3}}
\oplus
\overline{\mathbf{3}}_{-\frac{2}{3}}
\oplus
\mathbf{1}_{-1},
\end{eqnarray}
where the subscripts denote electric charge. These are precisely the electrocolour quantum numbers of one generation of leptons and quarks together with their charge-conjugate states. No weak-isospin or Lorentz-chirality assignment is implied at this stage.

An important feature of the construction is that gauge transformations act naturally through commutators in the ambient Clifford algebra. For $H\in\mathfrak{u}(3)$ and $X\in\mathcal I$, one has $fH=0$, and therefore
\begin{eqnarray}
[H,X]&=&HX.
\end{eqnarray}
Similarly, for $Y\in\mathcal I^c$ the conjugate generators satisfy
\begin{eqnarray}
[-H^\ast,Y]&=&-H^\ast Y.
\end{eqnarray}
Thus the adjoint action of the gauge algebra localises to the usual one-sided module action on the two selected fermion ideals.

\subsection{Why the sedenion step is useful}
\label{sec:why-sedenions-useful}

The passage from $\bb{O}$ to $\bb{S}$ sacrifices both the division property and alternativity. These losses do not, however, prevent the sedenions from playing a useful role in the present construction. The fermionic projector structure is not sought directly inside the real algebra $\bb{S}$. Indeed, although $\bb{S}$ contains zero divisors, it has no nontrivial idempotents \cite{BresarSemrlSpenko2011LocallyComplex}, so its zero divisors should not be confused with the primitive idempotents used later to construct algebraic spinors.

Nor is the loss of the division property unusual in algebraic models of particle physics. As noted above, complexified and tensor-product algebras such as $\bb{C}\otimes\bb{H}$ and $\bb{C}\otimes\bb{O}$ are themselves not division algebras and admit nontrivial projectors. In the octonion one-generation construction reviewed in the preceding subsection, the relevant primitive idempotents, minimal ideals, fermion states, and gauge generators are constructed not in the real division algebra $\bb{O}$, but in its associated complex Clifford multiplication algebra $\Cl{6}$.

There is also an important dimensional distinction between the octonion and sedenion cases. The complex octonions $\bb{C}\otimes\bb{O}$ have the same complex dimension as a minimal left ideal of $\Cl{6}$, so they may themselves be viewed as a spinor module. Likewise, $\bb{C}\otimes\bb{S}$ has complex dimension $16$, equal to that of a minimal left ideal of $\Cl{8}$, but this is sufficient for only one generation-sized spinor sector, not three linearly independent generations. A genuine three-generation construction therefore requires the larger multiplication algebra $\Cl{8}$ and its multiple spinor subspaces, rather than treating $\bb{C}\otimes\bb{S}$ itself as the complete fermion space.

The sedenion construction follows the same principle. The role of $\bb{S}$ is to supply two pieces of algebraic structure: an intrinsic $S_3$ automorphism symmetry and a natural route, through complex left multiplication, to $\Cl{8}$. The fermionic representation theory is then carried out in the associative Clifford algebra rather than in the raw nonassociative sedenion algebra. In the refined construction, $\Cl{8}$ becomes the common algebraic arena in which fermion states are represented by minimal ideals and semi-spinor subspaces, gauge symmetries by distinguished generators acting through commutators, and the $S_3$ family symmetry by automorphisms of the same structure.

The transition from the octonion to the sedenion construction may therefore be summarised schematically as
\begin{eqnarray}
\bb{C}\otimes\bb{O}
&\longrightarrow&
\Cl{6},
\\
\bb{C}\otimes\bb{S}
&\longrightarrow&
\Cl{8},
\end{eqnarray}
with the second case adding an intrinsic $S_3$ structure absent from the octonion one-generation construction. We now turn to the explicit realisation of this $\Cl{8}$ action algebra.

\subsection{Explicit realisation of $\Cl{8}$ from sedenion left actions}
\label{sec:CS-to-Cl8}

As established in Section~\ref{sec:multiplication-algebras}, the complex left multiplication algebra associated with $\bb{C}\otimes\bb{S}$ is isomorphic to $\Cl{8}$. We now make this identification explicit in the sedenion basis used throughout the review. Eight distinguished left multiplication maps provide Clifford generators, while the remaining sedenion directions are represented within the same algebra by higher-grade Clifford multivectors. This explicit correspondence is needed subsequently to realise the intrinsic sedenion $S_3$ automorphisms as transformations of $\Cl{8}$.

\subsubsection{Eight Clifford generators from left multiplication}

Let
\begin{eqnarray}
e_i := L_{s_i},\qquad i=1,\ldots,8.
\end{eqnarray}
Here the symbol $e_i$ denotes a Clifford generator, realised as a left multiplication map. It should not be confused with the sedenion basis element $s_i$ itself. With this convention, the left actions associated with $s_1,\ldots,s_8$ satisfy
\begin{eqnarray}
e_i^2 &=& -\mathbf{1}_{16},\qquad i=1,\ldots,8, \\
e_i e_j + e_j e_i &=& 0,\qquad i\neq j.
\end{eqnarray}
Equivalently,
\begin{eqnarray}
\{e_i,e_j\} &=& -2\delta_{ij}\mathbf{1}_{16}.
\label{eq:Cl8-generators}
\end{eqnarray}
These are precisely the defining relations of the Clifford algebra $C\ell(0,8)$. Once we complexify, the signature is immaterial. Thus, over $\bb{C}$, the algebra generated by the $e_i$ is the complex Clifford algebra $\Cl{8}$. That is, each (complex) sedenion left action corresponds to an element of $\textrm{Mat}(16,\bb{C})$, and all possible maps generate $\textrm{Mat}(16,\bb{C})\cong \Cl{8}$.

The complex dimension of $\Cl{8}$ is $2^8=256$. The full algebra of complex-linear endomorphisms of the $16$-dimensional vector space $\bb{C}\otimes \bb{S}$ also has complex dimension $16^2=256$. Thus the left multiplication algebra generated by these eight Clifford generators gives a faithful matrix representation
\begin{eqnarray}
\Cl{8} &\simeq& \operatorname{End}_{\bb{C}}(\bb{C}\otimes \bb{S}) \nonumber\\
&\simeq& \operatorname{Mat}(16,\bb{C}).
\label{eq:Cl8-Mat16}
\end{eqnarray}
This is the sense in which $\bb{C}\otimes \bb{S}$ gives rise to $\Cl{8}$.

\subsubsection{The remaining sedenion directions as Clifford multivectors}
\label{subsubsec:remainingsedeniondirections}

The eight generators $e_1,\ldots,e_8$ are sufficient to generate the full algebra $\Cl{8}$. The remaining left actions associated with $s_9,\ldots,s_{15}$ are therefore not independent generators of a larger Clifford algebra. Instead, they must correspond to certain multivectors in $\Cl{8}$.

It is useful to denote these remaining left actions by
\begin{eqnarray}
g_i := L_{s_{i+8}},\qquad i=1,\ldots,7.
\end{eqnarray}

In the chosen Clifford basis $\{e_i\}$, direct evaluation of the remaining sedenion left multiplication maps gives the following multivector representatives:
\begin{eqnarray}
g_1 &=& \frac{1}{2}e_1e_8(-e_{2345}+e_{2367}+e_{4567}-1), \\
g_2 &=& \frac{1}{2}e_2e_8(e_{1346}+e_{1357}+e_{4567}-1), \\
g_3 &=& \frac{1}{2}e_3e_8(e_{1256}-e_{1247}+e_{4567}-1), \\
g_4 &=& \frac{1}{2}e_4e_8(e_{1256}+e_{1357}+e_{2367}-1), \\
g_5 &=& \frac{1}{2}e_5e_8(-e_{1247}+e_{1346}+e_{2367}-1), \\
g_6 &=& \frac{1}{2}e_6e_8(-e_{1247}+e_{1357}-e_{2345}-1), \\
g_7 &=& \frac{1}{2}e_7e_8(e_{1256}+e_{1346}-e_{2345}-1),
\label{eq:gi-explicit}
\end{eqnarray}
where we use the shorthand
\begin{eqnarray}
e_{i_1i_2\cdots i_k}
&:=&
e_{i_1}e_{i_2}\cdots e_{i_k},
\end{eqnarray}
for ordered Clifford products. Thus, for example, $e_{2345}=e_2e_3e_4e_5$. These formulae can be written more compactly by introducing the four-vector combinations
\begin{eqnarray}
B_1 &=& -e_{2345}+e_{2367}+e_{4567}, \\
B_2 &=& e_{1346}+e_{1357}+e_{4567}, \\
B_3 &=& e_{1256}-e_{1247}+e_{4567}, \\
B_4 &=& e_{1256}+e_{1357}+e_{2367}, \\
B_5 &=& -e_{1247}+e_{1346}+e_{2367}, \\
B_6 &=& -e_{1247}+e_{1357}-e_{2345}, \\
B_7 &=& e_{1256}+e_{1346}-e_{2345}.
\label{eq:Bi-definition}
\end{eqnarray}
The structure of these four-vectors is inherited directly from the octonion $G_2$ structure introduced above. The same structure constants $\varphi_{ijk}$ defined in Eq.~\eqref{eq:octonion-multiplication} determine the $G_2$-invariant trivector
\begin{eqnarray}
\varphi
&=&
e_{123}+e_{145}-e_{167}+e_{246}+e_{257}+e_{347}-e_{356},
\label{eq:G2-three-form}
\end{eqnarray}
whose Hodge dual in the oriented seven-dimensional space spanned by $e_1,\ldots,e_7$ is
\begin{eqnarray}
\star\varphi
&=&
-e_{2345}+e_{2367}+e_{4567}
+e_{1346}+e_{1357}
+e_{1256}-e_{1247}.
\label{eq:G2-four-form}
\end{eqnarray}
Each $B_i$ in Eq.~\eqref{eq:Bi-definition} is obtained by retaining from $\star\varphi$ precisely those terms that do not contain the Clifford generator $e_i$. Equivalently, if $w := \varphi e_{1234567}$, then $w$ satisfies the quadratic identity $w^2 = 7+6w$. Thus the multivector representatives of the remaining sedenion left actions are controlled by the same $G_2$-invariant calibration structure that encodes the octonion multiplication.

Then the seven elements $B_i$ may be written uniformly as
\begin{eqnarray}
B_i&=&\sum_{1\leq j<k\leq 7}\varphi_{ijk}e_ie_je_ke_{1234567},\qquad i=1,\ldots,7,
\label{eq:Bi-G2-form}
\end{eqnarray}
with $e_{1234567}:=e_1e_2e_3e_4e_5e_6e_7$. Thus
\begin{eqnarray}
g_i&=&\frac{1}{2}e_i e_8(B_i-1),\qquad i=1,\ldots,7.
\label{eq:gi-definition}
\end{eqnarray}
The elements $g_i$ are therefore not additional Clifford generators. They are distinguished multivectors already contained in the algebra generated by $e_1,\ldots,e_8$.

Moreover, the seven elements $g_i$, together with $e_8$, form an alternative Clifford generating basis for the same algebra. The two sets
\begin{eqnarray}
\{e_1,\ldots,e_7,e_8\},
\qquad
\{g_1,\ldots,g_7,e_8\}
\end{eqnarray}
therefore provide two Clifford bases related by the sedenion structure underlying the order-three automorphism. In addition, each $g_i$ anticommutes with the corresponding original generator,
\begin{eqnarray}
e_i g_i+g_i e_i
&=&
0,
\qquad
i=1,\ldots,7.
\end{eqnarray}
However, the $g_i$ do not form a simple additional Clifford basis relative to all the $e_j$. For distinct indices $i\neq j$, one has in general
\begin{eqnarray}
\{g_i,e_j\}\neq0,\qquad [g_i,e_j]\neq0,\qquad i\neq j,\quad i,j=1,\ldots,7.
\end{eqnarray}
This makes the embedding of the sedenion automorphisms into $\Cl{8}$ nontrivial: the additional sedenion directions are represented by distinguished multivectors rather than by a second Clifford basis with simple mixed anticommutation relations.

\subsection{Matrix realisation, Witt basis, and minimal ideals}
\label{subsec:Cl8-witt-minimal-ideals}

Since $\Cl{8}\simeq\operatorname{Mat}(16,\bb{C})$, all faithful irreducible matrix realisations are related by similarity transformations, so the particular choice of basis is a matter of convenience rather than principle. We choose a realisation adapted to the Witt basis introduced below, for which the associated primitive idempotents are diagonal and the corresponding minimal left ideals are particularly transparent. Specifically,
\begin{eqnarray}
e_1 &=& \ii\sigma_1\otimes\sigma_1\otimes\sigma_1\otimes\sigma_1, \nonumber\\
e_2 &=& \ii\sigma_1\otimes\sigma_1\otimes\sigma_3\otimes\mathbf{1}_2, \nonumber\\
e_3 &=& -\ii\sigma_1\otimes\sigma_1\otimes\sigma_1\otimes\sigma_3, \nonumber\\
e_4 &=& -\ii\sigma_1\otimes\sigma_3\otimes\mathbf{1}_2\otimes\mathbf{1}_2, \nonumber\\
e_5 &=& -\ii\sigma_1\otimes\sigma_1\otimes\sigma_1\otimes\sigma_2, \nonumber\\
e_6 &=& -\ii\sigma_1\otimes\sigma_2\otimes\mathbf{1}_2\otimes\mathbf{1}_2, \nonumber\\
e_7 &=& \ii\sigma_1\otimes\sigma_1\otimes\sigma_2\otimes\mathbf{1}_2, \nonumber\\
e_8 &=& \ii\sigma_2\otimes\mathbf{1}_2\otimes\mathbf{1}_2\otimes\mathbf{1}_2,
\label{eq:Cl8-matrix-basis}
\end{eqnarray}
where $\sigma_i$, $i=1,2,3$ are the usual Pauli matrices.

These satisfy
\begin{eqnarray}
\{e_i,e_j\}
&=&
-2\delta_{ij}\mathbf{1}_{16}.
\end{eqnarray}
The corresponding pseudoscalar is
\begin{eqnarray}
\omega_8
&:=&
e_1e_2e_3e_4e_5e_6e_7e_8
=
\sigma_3\otimes\mathbf{1}_2\otimes\mathbf{1}_2\otimes\mathbf{1}_2,
\label{eq:omega8}
\end{eqnarray}
with
\begin{eqnarray}
\omega_8^2&=&\mathbf{1}_{16},
\qquad
\{\omega_8,e_i\}=0.
\end{eqnarray}

The associated four-pair Witt basis is
\begin{eqnarray}
a_j
&=&
\frac{1}{2}\left(-e_j+\ii e_{j+4}\right),
\qquad
a_j^\dagger
=
\frac{1}{2}\left(e_j+\ii e_{j+4}\right),
\qquad
j=1,\ldots,4,
\label{eq:Cl8-Witt-basis}
\end{eqnarray}
and satisfies
\begin{eqnarray}
\{a_i,a_j\}
=
\{a_i^\dagger,a_j^\dagger\}
&=&
0,
\qquad
\{a_i,a_j^\dagger\}
=
\delta_{ij}.
\label{eq:Cl8-Witt-relations}
\end{eqnarray}
As in the $\Cl{6}$ construction, the operators span complementary maximal totally isotropic subspaces
\begin{eqnarray}
W
&:=&
\operatorname{span}_{\bb{C}}\{a_1,a_2,a_3,a_4\},
\qquad
W^\dagger
:=
\operatorname{span}_{\bb{C}}\{a_1^\dagger,a_2^\dagger,a_3^\dagger,a_4^\dagger\}.
\end{eqnarray}

For each Witt pair define
\begin{eqnarray}
\pi_j^{(+)}
:=
a_j a_j^\dagger
=
\frac{1}{2}\left(1-\ii e_j e_{j+4}\right),
\qquad
\pi_j^{(-)}
:=
a_j^\dagger a_j
=
\frac{1}{2}\left(1+\ii e_j e_{j+4}\right).
\label{eq:Cl8-simple-idempotents}
\end{eqnarray}
The four pairs commute with one another, so choosing one idempotent from each pair gives the sixteen primitive idempotents
\begin{eqnarray}
f_{\epsilon_1\epsilon_2\epsilon_3\epsilon_4}
&:=&
\pi_1^{(\epsilon_1)}
\pi_2^{(\epsilon_2)}
\pi_3^{(\epsilon_3)}
\pi_4^{(\epsilon_4)},
\qquad
\epsilon_j\in\{+,-\}.
\label{eq:Cl8-primitive-idempotents}
\end{eqnarray}
They are mutually orthogonal and complete:
\begin{eqnarray}
f_\epsilon f_{\epsilon'}
&=&
\delta_{\epsilon,\epsilon'}f_\epsilon,
\qquad
\sum_{\epsilon\in\{+,-\}^4}f_\epsilon
=
\mathbf{1}_{16}.
\end{eqnarray}

In the chosen matrix realisation these are rank-one diagonal projectors. We denote by $E_{ij}$ the standard matrix units in $\operatorname{Mat}(16,\bb{C})$, defined by
\begin{eqnarray}
(E_{ij})_{kl}
&=&
\delta_{ik}\delta_{jl}.
\end{eqnarray}
Thus each $f_\epsilon$ is one of the diagonal matrix units $E_{ii}$.

The corresponding minimal left ideals are
\begin{eqnarray}
I_\epsilon
&:=&
\Cl{8}f_\epsilon,
\end{eqnarray}
each of complex dimension sixteen. In particular,
\begin{eqnarray}
f_{++++}
&=&
a_1a_1^\dagger
a_2a_2^\dagger
a_3a_3^\dagger
a_4a_4^\dagger
=
E_{11},
\end{eqnarray}
and
\begin{eqnarray}
I_1
&:=&
\Cl{8}f_{++++}.
\end{eqnarray}
By the same exterior-algebra construction reviewed for $\Cl{6}$,
\begin{eqnarray}
I_1
&\simeq&
\Lambda^\bullet W^\dagger f_{++++}.
\end{eqnarray}
Explicitly,
\begin{eqnarray}
I_1
&=&
\Big(
r_0
+r_i a_i^\dagger
+r_{ij}a_i^\dagger a_j^\dagger
+r_{ijk}a_i^\dagger a_j^\dagger a_k^\dagger
+r_{1234}a_1^\dagger a_2^\dagger a_3^\dagger a_4^\dagger
\Big)f_{++++},
\label{eq:I1-Cl8}
\end{eqnarray}
where the coefficients are complex and repeated indices are summed with $i<j$ and $i<j<k$ where appropriate. In the matrix realisation, $I_1=\Cl{8}E_{11}$ is the first column of $\operatorname{Mat}(16,\bb{C})$.

\section{Embedding the $S_3$ family action in $\Cl{8}$}
\label{sec:S3-in-Cl8}

We now describe how an $S_3$ action motivated by the intrinsic automorphism structure of the sedenions is realised inside $\Cl{8}$. The previous section showed that the complex sedenion left actions generate $\Cl{8}$ and that the remaining sedenion directions are represented in this algebra by the distinguished multivectors $g_i$. The sedenion automorphisms may therefore be transported to transformations of the Clifford generators. If the transformed generators again satisfy the Clifford relations, the resulting map extends to an automorphism of the full algebra $\Cl{8}$.

The order-three automorphism used below is denoted by $\psi_3$, while the order-two automorphism is denoted by $\epsilon$. Together they generate the $S_3$ action that will subsequently relate the three fermion sectors.

\subsection{Selection criteria for the embedded $S_3$ action}

The intrinsic order-three automorphism of $\bb{S}$ does not determine a unique physically useful action on $\Cl{8}$. The reason is that the passage from sedenion elements to their left multiplication operators is not an algebra homomorphism: in general,
\begin{eqnarray}
L_aL_b&\neq&L_{ab}.
\end{eqnarray}
Consequently, expressions that are equivalent inside the nonassociative algebra $\bb{S}$ can lead to distinct transformations after being translated into the associative Clifford multiplication algebra.

The simplest possibility is to translate the sedenion action in Eq.~\eqref{eq:sedenion-order-three-action} directly. Since
\begin{eqnarray}
e_i=L_{s_i},
\qquad
g_i=L_{s_{i+8}},
\end{eqnarray}
one may define
\begin{eqnarray}
\psi(e_i)
&:=&
-\frac{1}{2}e_i-\frac{\sqrt{3}}{2}g_i,
\qquad
i=1,\ldots,7,
\\
\psi(e_8)
&:=&
e_8.
\label{eq:direct-psi-embedding}
\end{eqnarray}
The induced action on the remaining sedenion left actions is
\begin{eqnarray}
\psi(g_i)
&=&
-\frac{1}{2}g_i+\frac{\sqrt{3}}{2}e_i.
\end{eqnarray}
This is a valid order-three automorphism of $\Cl{8}$ and preserves the inherited $G_2$ structure, including the colour algebra $\su(3)_C\subset\mathfrak{g}_2$. It is therefore the most direct Clifford realisation of the sedenion automorphism.

It is nevertheless insufficient for the three-generation construction. The orbit of the Clifford generators satisfies
\begin{eqnarray}
e_i+\psi(e_i)+\psi^2(e_i)
&=&
0,
\qquad
i=1,\ldots,7,
\end{eqnarray}
and the corresponding fermion sectors generated from this action are not fully linearly independent.

There is a second requirement coming from the gauge structure. In the $\Cl{6}$ one-generation construction reviewed above, preservation of the Witt decomposition selects
\begin{eqnarray}
\mathfrak{u}(3)
&=&
\mathfrak{su}(3)_C\oplus\mathfrak{u}(1)_{\rm em}.
\end{eqnarray}
A three-generation extension should retain a single corresponding electrocolour gauge algebra acting identically on all three family sectors. The colour part is naturally compatible with the sedenion symmetry because
\begin{eqnarray}
\su(3)_C
&\subset&
\mathfrak{g}_2,
\end{eqnarray}
and the intrinsic $S_3$ commutes with $G_2$. The abelian charge generator requires an additional condition, however, since the one-generation $\mathfrak{u}(1)_{\rm em}$ direction is not contained in $\mathfrak{g}_2$. The chosen order-three action must therefore admit an $S_3$-invariant abelian generator that reproduces the same electric-charge assignments in all three sectors. As shown later, this is achieved by combining the generation-resolved charge operators over their $S_3$ orbit.

The useful family action is therefore constrained simultaneously by several requirements. It must extend to an order-three automorphism of $\Cl{8}$, preserve the inherited $G_2$ structure and hence the single colour algebra, admit a family-invariant electromagnetic generator with the required charge spectrum, and generate three linearly independent fermion sectors. The direct embedding above satisfies the first two requirements but fails the latter conditions. More general Clifford realisations of the same underlying sedenion symmetry may then be considered, leading to the order-three map $\psi_3$ used in the remainder of the construction.

\subsection{The order-three generator $\psi_3$}

Having established the criteria that the family action must satisfy, we now introduce the order-three Clifford automorphism used in the remainder of the construction. Unlike the direct lift considered above, the selected map mixes the two distinguished sedenion left-action directions represented by $e_i$ and $g_i$, together with their products by $e_8$. It is defined on the Clifford generators by
\begin{eqnarray}
\psi_3(e_i)
&=&
\frac{1}{4}e_i
+\frac{\sqrt{3}}{4}g_i
-\frac{\sqrt{3}}{4}e_i e_8
-\frac{3}{4}g_i e_8,
\qquad
i=1,\ldots,7,
\label{eq:psi3-on-ei}
\end{eqnarray}
and
\begin{eqnarray}
\psi_3(e_8)
&=&
e_8.
\label{eq:psi3-on-e8}
\end{eqnarray}

The transformed generators satisfy the same Clifford relations as the original generating basis,
\begin{eqnarray}
\{\psi_3(e_i),\psi_3(e_j)\}
&=&
-2\delta_{ij}\mathbf{1}_{16},
\qquad
i,j=1,\ldots,8.
\label{eq:psi3-Clifford-relations}
\end{eqnarray}
Thus $\psi_3(e_1),\ldots,\psi_3(e_8)$ constitute another Clifford generating basis, and the assignment $e_i\mapsto\psi_3(e_i)$ extends uniquely to an algebra automorphism of $\Cl{8}$.

Since the multivectors $g_i$ are themselves elements of $\Cl{8}$ defined in terms of the original generators, their transformation is fixed by this automorphism. One finds
\begin{eqnarray}
\psi_3(g_i)
&=&
-\frac{\sqrt{3}}{4}e_i
+\frac{1}{4}g_i
+\frac{3}{4}e_i e_8
-\frac{\sqrt{3}}{4}g_i e_8,
\qquad
i=1,\ldots,7.
\label{eq:psi3-on-gi}
\end{eqnarray}
Applying the automorphism a second time gives
\begin{eqnarray}
\psi_3^2(e_i)
&=&
\frac{1}{4}e_i
-\frac{\sqrt{3}}{4}g_i
+\frac{\sqrt{3}}{4}e_i e_8
-\frac{3}{4}g_i e_8,
\qquad
i=1,\ldots,7.
\label{eq:psi3-squared-on-ei}
\end{eqnarray}
A third application returns every Clifford generator to itself,
\begin{eqnarray}
\psi_3^3(e_i)
&=&
e_i,
\qquad
i=1,\ldots,8,
\end{eqnarray}
and hence, since the $e_i$ generate the full algebra,
\begin{eqnarray}
\psi_3^3
&=&
\operatorname{Id}.
\label{eq:psi3-order-three}
\end{eqnarray}
Therefore $\psi_3$ generates the cyclic subgroup
\begin{eqnarray}
\langle\psi_3\rangle
&\simeq&
\bb{Z}_3
\subset
S_3,
\end{eqnarray}
which will subsequently generate the three fermion sectors.

It is important that $\psi_3$ is not a $\Spin(8)$ transformation. The adjoint action of $\Spin(8)$ preserves the canonical one-vector subspace, whereas $\psi_3(e_i)$ contains components of different Clifford grades. In particular, because $g_i$ and $e_i e_8$ are even while $e_i$ and $g_i e_8$ are odd, $\psi_3$ does not preserve the natural $\bb{Z}_2$ grading of $\Cl{8}$. It is therefore an automorphism of the full Clifford algebra, but not an element of the $\Spin(8)$ action associated with the original one-vector space.

\subsection{The order-two automorphism and the $S_3$ relations}

As for the order-three generator, the most direct possibility is to translate the corresponding sedenion automorphism into the Clifford action algebra. The sedenion order-two automorphism introduced in Eq.~\eqref{eq:sedenion-order-two-action} acts by reversing the sign of the component proportional to $s_8$. Since
\begin{eqnarray}
e_i=L_{s_i},
\qquad
i=1,\ldots,8,
\end{eqnarray}
its direct Clifford translation is
\begin{eqnarray}
\widetilde{\epsilon}(e_i)
&=&
e_i,
\qquad
i=1,\ldots,7,
\\
\widetilde{\epsilon}(e_8)
&=&
-e_8.
\end{eqnarray}
and consequently
\begin{eqnarray}
\widetilde{\epsilon}(g_i)
&=&
-g_i.
\end{eqnarray}
In the Witt basis it leaves the first three pairs unchanged but interchanges the fourth creation and annihilation operators. It therefore exchanges
\begin{eqnarray}
\widetilde{\epsilon}\left(\pi_4^{(+)}\right)
&=&
\pi_4^{(-)},
\qquad
\widetilde{\epsilon}\left(\pi_4^{(-)}\right)
=
\pi_4^{(+)},
\end{eqnarray}
and moves between distinct primitive-idempotent sectors. Although this action preserves the colour algebra, it does not leave invariant the $U(1)$ generator introduced later and, upon extension to $\Cl{10}$, does not preserve the weak $\SU(2)_L$ gauge generators.

It is therefore more useful to identify the order-two automorphism of $S_3$ with the grade involution of $\Cl{8}$,
\begin{eqnarray}
\epsilon(e_i)
&=&
-e_i,
\qquad
i=1,\ldots,8.
\end{eqnarray}
Since the elements $g_i$ are even multivectors, this gives
\begin{eqnarray}
\epsilon(g_i)
&=&
g_i,
\qquad
i=1,\ldots,7.
\end{eqnarray}
Equivalently, the Witt generators transform as
\begin{eqnarray}
\epsilon(a_j)
&=&
-a_j,
\qquad
\epsilon(a_j^\dagger)
=
-a_j^\dagger,
\qquad
j=1,\ldots,4.
\end{eqnarray}
All quadratic Witt bilinears are therefore fixed, and in particular
\begin{eqnarray}
\epsilon\left(\pi_j^{(\pm)}\right)
&=&
\pi_j^{(\pm)},
\qquad
j=1,\ldots,4.
\end{eqnarray}
Consequently every primitive idempotent $f_{\epsilon_1\epsilon_2\epsilon_3\epsilon_4}$ is fixed by $\epsilon$. This choice is compatible with the requirement that the gauge generators be invariant under the family symmetry.

Together, $\psi_3$ and $\epsilon$ satisfy
\begin{eqnarray}
\psi_3^3
&=&
\operatorname{Id},
\qquad
\epsilon^2
=
\operatorname{Id},
\qquad
\epsilon\psi_3
=
\psi_3^2\epsilon.
\label{eq:S3-relations-Cl8}
\end{eqnarray}
Equivalently,
\begin{eqnarray}
\epsilon\psi_3\epsilon
&=&
\psi_3^{-1}.
\end{eqnarray}
Therefore
\begin{eqnarray}
\left\langle
\psi_3,\epsilon
\ \big|\
\psi_3^3=\operatorname{Id},\
\epsilon^2=\operatorname{Id},\
\epsilon\psi_3=\psi_3^2\epsilon
\right\rangle
&\simeq&
S_3.
\end{eqnarray}
The cyclic subgroup generated by $\psi_3$ produces the three family sectors, while the full $S_3$ action will become important when the family representation is decomposed into its singlet and doublet components.

\subsection{Action of $\psi_3$ on primitive idempotents}

We next record the action of $\psi_3$ on the primitive idempotents that will later support the fermion subspaces. The full algebra $\Cl{8}$ contains sixteen primitive idempotents of the form $f_{\epsilon_1\epsilon_2\epsilon_3\epsilon_4}$, but the four idempotents displayed below are the ones relevant for the closed $S_3$ subspaces and for the colour-compatible semi-spinor sectors used in the three-generation construction. In the matrix realisation used here, they are
\begin{eqnarray}
f_{++++} = E_{1,1}, \qquad
f_{---+} = E_{14,14}, \qquad
f_{+++-} = E_{9,9}, \qquad
f_{----} = E_{6,6}.
\end{eqnarray}
The action of $\psi_3$ on these primitive idempotents is not simply a permutation of the original diagonal idempotents. Although $\psi_3(f)$ is again a primitive idempotent, its expression in the original matrix-unit basis is generally a nontrivial linear combination of matrix units. A direct calculation gives
\begin{eqnarray}
\psi_3(E_{1,1})&=&\frac{1}{4}E_{1,1}+\frac{3}{4}E_{14,14}+\frac{\ii\sqrt{3}}{4}\left(E_{14,1}-E_{1,14}\right),\label{eq:psi3-E11}\\
\psi_3^2(E_{1,1})&=&\frac{1}{4}E_{1,1}+\frac{3}{4}E_{14,14}-\frac{\ii\sqrt{3}}{4}\left(E_{14,1}-E_{1,14}\right).
\label{eq:psi3sq-E11}
\end{eqnarray}
Acting once more returns $E_{1,1}$, as required. Similarly,
\begin{eqnarray}
\psi_3(E_{14,14})&=&\frac{3}{4}E_{1,1}+\frac{1}{4}E_{14,14}+\frac{\ii\sqrt{3}}{4}\left(E_{1,14}-E_{14,1}\right),
\\
\psi_3^2(E_{14,14})&=&\frac{3}{4}E_{1,1}+\frac{1}{4}E_{14,14}-\frac{\ii\sqrt{3}}{4}\left(E_{1,14}-E_{14,1}\right).
\end{eqnarray}
It follows that the orbit average
\begin{eqnarray}
E_{1,1}+\psi_3(E_{1,1})+\psi_3^2(E_{1,1})&=&\frac{3}{2}\left(E_{1,1}+E_{14,14}\right),
\end{eqnarray}
is $S_3$-invariant.

The same structure appears for the pair $E_{9,9}$ and $E_{6,6}$. One finds
\begin{eqnarray}
\psi_3(E_{9,9})&=&\frac{1}{4}E_{9,9}+\frac{3}{4}E_{6,6}+\frac{\ii\sqrt{3}}{4}\left(E_{9,6}-E_{6,9}\right),
\\
\psi_3^2(E_{9,9})&=&\frac{1}{4}E_{9,9}+\frac{3}{4}E_{6,6}-\frac{\ii\sqrt{3}}{4}\left(E_{9,6}-E_{6,9}\right),
\end{eqnarray}
and hence
\begin{eqnarray}
E_{9,9}+\psi_3(E_{9,9})+\psi_3^2(E_{9,9})&=&\frac{3}{2}\left(E_{9,9}+E_{6,6}\right).
\end{eqnarray}
These formulae show explicitly how $\psi_3$ moves primitive idempotent sectors into nontrivial linear combinations. This is why the $S_3$ family action is not simply a permutation of the original matrix-unit idempotents.

\subsection{$S_3$-closed spinor subspaces}
\label{subsec:S3-closed-spinor-subspaces}

The preceding formulae imply that certain sums of minimal left ideals are closed under the $S_3$ action. In particular,
\begin{eqnarray}
\Cl{8}f_{++++}
\oplus
\Cl{8}f_{---+},
\end{eqnarray}
is closed under both $\psi_3$ and $\epsilon$. Likewise,
\begin{eqnarray}
\Cl{8}f_{+++-}
\oplus
\Cl{8}f_{----},
\end{eqnarray}
is closed under both $\psi_3$ and $\epsilon$. More generally, the sixteen primitive idempotents split into eight two-idempotent blocks whose sums are fixed by $\psi_3$. The two blocks displayed here are the ones relevant for the later colour-compatible fermion subspaces; the remaining closed blocks are part of the ambient Clifford algebra but are not used in the fermion-sector construction below.

This closure property is essential for the later construction. It means that the embedded $S_3$ action does not take us out of the relevant closed block of Clifford spinor space. Instead, it acts internally on controlled finite-dimensional sums of minimal left ideals in $\Cl{8}$.

Finally, because $\psi_3$ and $\epsilon$ are algebra automorphisms, they preserve commutators. Thus, for any $X,Y\in\Cl{8}$,
\begin{eqnarray}
\psi_3([X,Y]) &=& [\psi_3(X),\psi_3(Y)], \\
\epsilon([X,Y]) &=& [\epsilon(X),\epsilon(Y)].
\end{eqnarray}
This covariance becomes particularly important for the gauge action. If a gauge generator $G$ is invariant under the family symmetry, then for any family transformation $\sigma\in S_3$,
\begin{eqnarray}
\sigma(G)
&=&
G,
\end{eqnarray}
implies
\begin{eqnarray}
\sigma([G,X])
&=&
[\sigma(G),\sigma(X)]
=
[G,\sigma(X)].
\label{eq:S3-gauge-covariance}
\end{eqnarray}
Consequently, states related by the $S_3$ family action transform identically under any $S_3$-invariant gauge generator. This is the algebraic condition that allows the family symmetry to act on the fermion sectors while leaving a single gauge sector unreplicated.

\section{Three generations and an untriplicated $SU(3)_C\times U(1)_{\rm em}$ sector in $\Cl{8}$}
\label{sec:three-generations-Cl8}

The aim of this section is to show how the $S_3$ action introduced above produces three linearly independent generation sectors transforming under a single untriplicated gauge sector $SU(3)_C\times U(1)_{\rm em}$. This construction should be regarded as the unbroken-gauge precursor of the later $\bb{C}\ell(10)$ model, where the full electroweak structure is included.

The main idea is simple. One first constructs a fermion sector inside $\Cl{8}$ using semi-spinors obtained from minimal left ideals. The order-three generator $\psi_3$ then produces two further generation sectors. The gauge generators are chosen to be invariant under $\psi_3$, so the three generations transform equivalently under the same gauge group. Thus the fermion sector is triplicated, but the gauge sector is not.

\subsection{Colour symmetry and localisation of the commutator action}

As reviewed in Section~\ref{sec:Cl6-one-generation-template}, preservation of an $n$-pair Witt decomposition selects a $\mathfrak{u}(n)$ algebra of unitary internal transformations. For the four-pair Witt basis of $\Cl{8}$ this gives $\mathfrak{u}(4)$. The physical gauge algebra is more restricted, however: its generators must act appropriately on the selected fermion subspaces and must be invariant under the embedded $S_3$ family symmetry.

Within this $\mathfrak{u}(4)$, the first three Witt pairs define a distinguished $\mathfrak{u}(3)$ subalgebra. Its traceless part is identified with the colour algebra $\mathfrak{su}(3)_C$.

As in the $\Cl{6}$ one-generation construction, the colour algebra acts on the first three Witt pairs. Writing
\begin{eqnarray}
\mathbf{a}
&:=&
\left(a_1,a_2,a_3\right)^T,
\end{eqnarray}
the colour generators are
\begin{eqnarray}
\Lambda_a
&:=&
-\mathbf{a}^{\dagger}\lambda_a^T\mathbf{a},
\qquad
a=1,\ldots,8,
\label{eq:Cl8-su3-generators}
\end{eqnarray}
where $\lambda_a$ are the standard Gell--Mann matrices. These generators lie in the $\mathfrak{u}(3)\subset\mathfrak{u}(4)$ subalgebra acting on the first three Witt pairs.

These generators are fixed by the full family symmetry:
\begin{eqnarray}
\psi_3(\Lambda_a)
&=&
\Lambda_a,
\qquad
\epsilon(\Lambda_a)
=
\Lambda_a,
\qquad
a=1,\ldots,8.
\label{eq:Lambda-S3-invariant}
\end{eqnarray}
As discussed above, this reflects the inclusion $\mathfrak{su}(3)_C\subset\mathfrak{g}_2$ together with
\begin{eqnarray}
\Aut(\bb{S})
&\simeq&
G_2\times S_3,
\end{eqnarray}
so that the family symmetry commutes with the inherited colour structure. A single colour algebra therefore acts on all three fermion sectors.

The choice of fermion-supporting minimal ideals is further constrained by the requirement that the colour commutator action reduce to a left-module action. In the full Clifford algebra,
\begin{eqnarray}
[\Lambda_a,X]
&=&
\Lambda_aX-X\Lambda_a,
\end{eqnarray}
contains both left and right multiplication, so this localisation does not occur on every minimal ideal. In the matrix realisation used here, exactly four of the sixteen primitive idempotents have the required localisation property:
\begin{eqnarray}
f_{++++}=E_{1,1},\qquad
f_{---+}=E_{14,14},\qquad
f_{+++-}=E_{9,9},\qquad
f_{----}=E_{6,6}.
\end{eqnarray}
The corresponding minimal left ideals are
\begin{eqnarray}
I_1=\Cl{8}f_{++++},\qquad
I_{14}=\Cl{8}f_{---+},\qquad
I_9=\Cl{8}f_{+++-},\qquad
I_6=\Cl{8}f_{----}.
\end{eqnarray}
For the non-Cartan colour generators one has directly
\begin{eqnarray}
[\Lambda_a,X]&=&\Lambda_aX,\qquad
a\in\{1,2,4,5,6,7\},
\qquad X\in I_k,
\qquad k\in\{1,6,9,14\}.
\end{eqnarray}
For the Cartan generators $\Lambda_3$ and $\Lambda_8$, the right action can be absorbed into an ideal-dependent representative differing from the original generator by an element acting trivially on the relevant ideal. Thus, on each of these four ideals, the full $\mathfrak{su}(3)_C$ commutator action is equivalent to a purely left action. For the remaining twelve primitive idempotents, at least one non-Cartan colour generator has unavoidable right-action support, so no such localisation occurs. These four ideals are precisely those from which the fermion semi-spinors used below are constructed.

It is important that commuting with the colour algebra is not by itself sufficient for family invariance. The diagonal $\mathfrak{su}(4)$ generator
\begin{eqnarray}
\Lambda_{15}&:=&-\frac{1}{\sqrt{6}}\left(a_1^\dagger a_1+a_2^\dagger a_2+a_3^\dagger a_3-3a_4^\dagger a_4,
\right)
\end{eqnarray}
satisfies
\begin{eqnarray}
[\Lambda_{15},\Lambda_i]&=&0,\qquad i=1,\ldots,8,
\end{eqnarray}
but it is not fixed by the order-three family action:
\begin{eqnarray}
\psi_3(\Lambda_{15})&\neq&\Lambda_{15}.
\end{eqnarray}
Indeed, using $a_j a_j^\dagger=1-a_j^\dagger a_j$, the one-generation charge operator introduced below satisfies
\begin{eqnarray}
\Lambda_{15}&=&\sqrt{\frac{3}{2}}Q_1.
\end{eqnarray}
The failure of $\Lambda_{15}$, or equivalently $Q_1$, to be $\psi_3$-invariant is the reason that the physical electromagnetic generator must instead be constructed by averaging over the order-three family orbit.

\subsection{The reference fermion sector from colour-compatible semi-spinors}

We now select the Clifford subspaces that will represent one fermion generation. The choice is constrained first by the colour action. As shown above, the $\mathfrak{su}(3)_C$ commutator action localises to a left action on only four of the sixteen minimal left ideals. We begin with the previously introduced vacuum ideal
\begin{eqnarray}
I_1
&:=&
\Cl{8}f_{++++},
\end{eqnarray}
where
\begin{eqnarray}
a_i f_{++++}
&=&
0,
\qquad
i=1,\ldots,4.
\end{eqnarray}
This is one of the four ideals compatible with the localised colour action.

A full minimal left ideal cannot, however, be identified with one generation. Under $\mathfrak{su}(3)_C$, the ideal $I_1$ decomposes as
\begin{eqnarray}
I_1
&\simeq&
\left(\mathbf{1}\oplus\mathbf{3}\oplus\overline{\mathbf{3}}\oplus\mathbf{1}\right)
\oplus
\left(\mathbf{1}\oplus\mathbf{3}\oplus\overline{\mathbf{3}}\oplus\mathbf{1}\right),
\end{eqnarray}
with the two copies carried by its even and odd semi-spinors. Moreover, as shown in Section~\ref{subsec:S3-closed-spinor-subspaces}, the $S_3$ orbit of $I_1$ is contained in the closed $32$-dimensional space
\begin{eqnarray}
I_1\oplus I_{14},
\end{eqnarray}
and therefore cannot contain three linearly independent $16$-dimensional spinors. This motivates passing to the $8$-dimensional semi-spinors of $I_1$.

Using the pseudoscalar $\omega_8$ introduced above, define the projectors
\begin{eqnarray}
\rho^\pm
&:=&
\frac{1}{2}(1\pm\omega_8),
\label{eq:rho-plus-minus-Cl8}
\end{eqnarray}
and hence
\begin{eqnarray}
I_1^+
&:=&
\rho^+I_1,
\qquad
I_1^-:=\rho^-I_1.
\end{eqnarray}
The even semi-spinor is
\begin{eqnarray}
I_1^+
&=&
\bb{C}\ell^+(8)f_{++++}
\\
&=&
\left(r_0+r_{ij}a_i^\dagger a_j^\dagger+r_{i4}a_i^\dagger a_4^\dagger
+r_{1234}a_1^\dagger a_2^\dagger a_3^\dagger a_4^\dagger\right)f_{++++},
\label{eq:I1-plus}
\end{eqnarray}
where $i,j\in\{1,2,3\}$ with $i<j$, and the coefficients are complex. Under $\mathfrak{su}(3)_C$,
\begin{eqnarray}
I_1^+
&\simeq&
\mathbf{1}\oplus\mathbf{3}\oplus\overline{\mathbf{3}}\oplus\mathbf{1}.
\label{eq:I1-plus-su3}
\end{eqnarray}

A second $8$-dimensional colour multiplet is required to complete the electrocolour content of one generation. Since the colour generators involve only the first three Witt pairs, the fourth annihilation operator is a colour singlet:
\begin{eqnarray}
[\Lambda_a,a_4]
&=&
0,
\qquad
a=1,\ldots,8.
\end{eqnarray}
Right multiplication by $a_4$ therefore preserves the colour representation and defines
\begin{eqnarray}
I_9^-
&:=&
I_1^+a_4.
\label{eq:I9-minus-def}
\end{eqnarray}
This is an odd semi-spinor built on
\begin{eqnarray}
f_{+++-}
&=&
a_1a_1^\dagger a_2a_2^\dagger a_3a_3^\dagger a_4^\dagger a_4,
\end{eqnarray}
so that
\begin{eqnarray}
I_9^-
&=&
\bb{C}\ell^-(8)f_{+++-}.
\end{eqnarray}
It transforms as
\begin{eqnarray}
I_9^-
&\simeq&
\mathbf{1}\oplus\mathbf{3}\oplus\overline{\mathbf{3}}\oplus\mathbf{1},
\end{eqnarray}
under $\mathfrak{su}(3)_C$.

The reference generation sector is therefore
\begin{eqnarray}
\Gen{1}
&:=&
I_1^+\oplus I_9^-.
\label{eq:first-generation-Cl8}
\end{eqnarray}
It is a $16$-complex-dimensional space consisting of two complementary $8$-dimensional semi-spinors. 

\subsection{Three linearly independent generation sectors}

The order-three automorphism generates two further fermion sectors from the reference sector:
\begin{eqnarray}
\Gen{2}
&:=&
\psi_3(\Gen{1}),
\qquad
\Gen{3}
:=
\psi_3^2(\Gen{1}).
\label{eq:three-generations-Cl8}
\end{eqnarray}
Since $\psi_3^3=\operatorname{Id}$, the orbit closes:
\begin{eqnarray}
\psi_3(\Gen{3})
&=&
\Gen{1}.
\end{eqnarray}
The three sectors are therefore not separate copies postulated independently. They form the orbit of a single fermion sector under the order-three family action.

The action on individual states is nontrivial. For example, in the matrix realisation used above,
\begin{eqnarray}
a_1^\dagger a_4^\dagger f_{++++}
&=&
E_{8,1},
\end{eqnarray}
and
\begin{eqnarray}
\psi_3\left(a_1^\dagger a_4^\dagger f_{++++}\right)
&=&
\frac{1}{4}E_{8,1}
-\frac{\ii\sqrt{3}}{4}E_{8,14}
-\frac{\sqrt{3}}{4}E_{16,1}
+\frac{3\ii}{4}E_{16,14}.
\label{eq:psi3-example-matrix}
\end{eqnarray}
Equivalently, in terms of the original Witt basis,
\begin{eqnarray}
\psi_3\left(a_1^\dagger a_4^\dagger f_{++++}\right)
=
\frac{1}{4}a_1^\dagger a_4^\dagger f_{++++}
+\frac{\ii\sqrt{3}}{4}a_1^\dagger f_{++++}
+\frac{3\ii}{4}a_2a_3f_{---+}
-\frac{\sqrt{3}}{4}a_2a_3a_4^\dagger f_{---+}.
\label{eq:psi3-example-Witt}
\end{eqnarray}
Thus the family action does not merely attach a generation label to an otherwise unchanged state. It maps a state in the reference sector to a specific linear combination of Clifford states within the relevant $S_3$-closed ideal support.

In the explicit matrix realisation, the combined basis states spanning the three sectors have rank $48$. Equivalently,
\begin{eqnarray}
\dim_{\bb{C}}\left(\Gen{1}+\Gen{2}+\Gen{3}\right)
&=&
48
\\
&=&
\dim_{\bb{C}}\Gen{1}
+\dim_{\bb{C}}\Gen{2}
+\dim_{\bb{C}}\Gen{3}.
\label{eq:Cl8-generation-rank}
\end{eqnarray}
Hence
\begin{eqnarray}
\Gen{1}+\Gen{2}+\Gen{3}
&=&
\Gen{1}\oplus\Gen{2}\oplus\Gen{3}.
\end{eqnarray}
This is the precise sense in which the construction produces three linearly independent generation sectors. As discussed above, the use of the selected semi-spinors is essential for this dimensional independence.

The total fermion space of the unbroken-gauge $\Cl{8}$ model is consequently
\begin{eqnarray}
\mathcal{F}_{\Cl{8}}
&:=&
\Gen{1}\oplus\Gen{2}\oplus\Gen{3}
\\
&=&
\left(I_1^+\oplus I_9^-\right)
\oplus
\psi_3\left(I_1^+\oplus I_9^-\right)
\oplus
\psi_3^2\left(I_1^+\oplus I_9^-\right).
\label{eq:F-Cl8}
\end{eqnarray}

Because the colour generators are fixed by $\psi_3$, the three sectors are equivalent as $\mathfrak{su}(3)_C$ modules:
\begin{eqnarray}
\Gen{1}
\simeq
\Gen{2}
\simeq
\Gen{3}.
\end{eqnarray}
The three fermion sectors are therefore linearly independent but gauge-equivalent under the single untriplicated colour algebra.

\subsection{The $S_3$-invariant electromagnetic charge}

In the $\Cl{6}$ one-generation construction reviewed above, the electromagnetic generator is obtained directly from the three-oscillator number operator,
\begin{eqnarray}
Q_{\Cl{6}}
&:=&
\frac{1}{3}\sum_{i=1}^{3}\alpha_i^\dagger\alpha_i.
\end{eqnarray}
This operator commutes with $\mathfrak{su}(3)_C$ and assigns the familiar electric charges to the states in the two selected conjugate minimal ideals.

The direct number-operator prescription does not extend to the selected $\Cl{8}$ fermion sector. Neither the three-mode operator
\begin{eqnarray}
N_3
&:=&
\sum_{i=1}^{3}a_i^\dagger a_i,
\end{eqnarray}
nor its four-mode analogue
\begin{eqnarray}
N_4
&:=&
\sum_{i=1}^{4}a_i^\dagger a_i,
\end{eqnarray}
reproduces the required electric-charge spectrum on $\Gen{1}=I_1^+\oplus I_9^-$. Moreover, neither operator is fixed by $\psi_3$, and although their order-three orbit averages are family-invariant, the resulting spectra remain incorrect. A different abelian direction is therefore required.

The centraliser of $\mathfrak{su}(3)_C$ inside $\mathfrak{su}(4)$ is generated by
\begin{eqnarray}
\Lambda_{15}
&:=&
-\frac{1}{\sqrt{6}}
\left(
a_1^\dagger a_1
+a_2^\dagger a_2
+a_3^\dagger a_3
-3a_4^\dagger a_4
\right).
\label{eq:Lambda15-Cl8}
\end{eqnarray}
A convenient generation-resolved abelian operator adapted to the reference Witt basis is
\begin{eqnarray}
Q_1
&:=&
\frac{1}{3}
\left(
a_1a_1^\dagger
+a_2a_2^\dagger
+a_3a_3^\dagger
-3a_4a_4^\dagger
\right)
=
\sqrt{\frac{2}{3}}\Lambda_{15}.
\label{eq:Q1-Cl8}
\end{eqnarray}
The final equality follows from $a_i a_i^\dagger=1-a_i^\dagger a_i$, and makes manifest that
\begin{eqnarray}
[Q_1,\Lambda_a]
&=&
0,
\qquad
a=1,\ldots,8.
\end{eqnarray}
Thus $Q_1$ is compatible with the colour algebra, but it is not invariant under the family action:
\begin{eqnarray}
\psi_3(Q_1)
&\neq&
Q_1.
\end{eqnarray}

The family-invariant electromagnetic generator is obtained by averaging $Q_1$ over its order-three orbit:
\begin{eqnarray}
Q
&:=&
\frac{1}{3}
\left(
Q_1+\psi_3(Q_1)+\psi_3^2(Q_1)
\right).
\label{eq:Q-S3-average}
\end{eqnarray}
It is fixed by the full family symmetry,
\begin{eqnarray}
\psi_3(Q)
&=&
Q,
\qquad
\epsilon(Q)
=
Q,
\end{eqnarray}
and commutes with the common colour algebra,
\begin{eqnarray}
[Q,\Lambda_a]
&=&
0,
\qquad
a=1,\ldots,8.
\end{eqnarray}
By the family--gauge covariance established in Eq.~\eqref{eq:S3-gauge-covariance}, the $S_3$ invariance of $Q$ ensures that corresponding states in the three generation sectors carry identical electric charges. The three operators $Q_1$, $\psi_3(Q_1)$, and $\psi_3^2(Q_1)$ are auxiliary generation-resolved elements; only their $S_3$-invariant average $Q$ is identified with the physical electromagnetic generator.

The averaging also changes the abelian spectrum. For example,
\begin{eqnarray}
[Q_1,a_i^\dagger a_j^\dagger f_{++++}]
&=&
-\frac{2}{3}a_i^\dagger a_j^\dagger f_{++++},
\\
{}[Q,a_i^\dagger a_j^\dagger f_{++++}]
&=&
\frac{1}{3}a_i^\dagger a_j^\dagger f_{++++}.
\end{eqnarray}

The charge eigenvalues on the reference fermion sector are summarised in Table~\ref{tab:Cl8-charge-table}. Together with the $\mathfrak{su}(3)_C$ representations, they reproduce the electrocolour multiplets of one fermion generation. The two neutral colour singlets have identical $\SUthreeC\times\Uoneem$ quantum numbers and therefore cannot yet be distinguished as neutrino-type and antineutrino-type states by the electrocolour symmetry alone. The same colour and charge assignments hold for the corresponding states in the two $\psi_3$-generated sectors. At this stage, the three algebraic sectors also cannot be identified physically with the electron, muon, and tau generations.

\begin{table}[h]
\centering
\begin{tabular}{c c c}
\hline
State & $Q$ eigenvalue & Interpretation \\
\hline
$f_{++++}$ & $0$ & neutrino-type singlet \\
$a_i^\dagger a_j^\dagger f_{++++}$ & $+\frac{1}{3}$ & anti-down-type colour states \\
$a_i^\dagger a_4^\dagger f_{++++}$ & $+\frac{2}{3}$ & up-type colour states \\
$a_1^\dagger a_2^\dagger a_3^\dagger a_4^\dagger f_{++++}$ & $+1$ & positron-type singlet \\
\hline
$a_4 f_{+++-}$ & $-1$ & electron-type singlet \\
$a_i^\dagger a_j^\dagger a_4 f_{+++-}$ & $-\frac{2}{3}$ & anti-up-type colour states \\
$a_i^\dagger f_{+++-}$ & $-\frac{1}{3}$ & down-type colour states \\
$a_1^\dagger a_2^\dagger a_3^\dagger f_{+++-}$ & $0$ & antineutrino-type singlet \\
\hline
\end{tabular}
\caption{Eigenvalues of the $S_3$-invariant electromagnetic generator $Q$ on the reference $\Cl{8}$ fermion sector $\Gen{1}$. Here $i,j\in\{1,2,3\}$ with $i<j$.}
\label{tab:Cl8-charge-table}
\end{table}

Unlike the nonabelian colour action, the electromagnetic commutator need not reduce to left multiplication by $Q$ itself. On the selected charge eigenstates, its physically relevant action is diagonal:
\begin{eqnarray}
[Q,X]
&=&
q_X X.
\end{eqnarray}
A one-sided description can nevertheless be given on each fixed minimal left ideal. If $X\in\Cl{8}f$ and
\begin{eqnarray}
fQ
&=&
\chi_f f,
\end{eqnarray}
then
\begin{eqnarray}
[Q,X]
&=&
\left(Q-\chi_f\mathbf{1}\right)X.
\end{eqnarray}
Thus the adjoint action may be represented on each ideal by an ideal-dependent shifted left generator, but there is no single universal left representative of $Q$ on the entire fermion space. For the abelian gauge factor, the representation is instead characterised directly by the charge eigenvalues and their $S_3$ covariance.

\subsection{Family representation and untriplicated gauge structure}

The order-three generator acts cyclically on the three generation sectors, while the order-two generator fixes the reference sector and exchanges its two images:
\begin{eqnarray}
\psi_3:\quad
\Gen{1}\longrightarrow\Gen{2}\longrightarrow\Gen{3}\longrightarrow\Gen{1},
\qquad
\epsilon(\Gen{1})=\Gen{1},
\qquad
\epsilon(\Gen{2})=\Gen{3},
\qquad
\epsilon(\Gen{3})=\Gen{2}.
\end{eqnarray}
Thus, at the level of the three generation sectors, $S_3$ acts through its natural permutation representation $\mathbf{1}\oplus\mathbf{2}$.

The action on individual states also reflects the even--odd decomposition of the reference sector. On the two semi-spinors,
\begin{eqnarray}
\epsilon\big|_{I_1^+}
&=&
+\operatorname{Id},
\qquad
\epsilon\big|_{I_9^-}
=
-\operatorname{Id}.
\end{eqnarray}
Consequently, each of the eight state directions in $I_1^+$ generates an $S_3$ orbit transforming as $\mathbf{1}\oplus\mathbf{2}$, whereas each of the eight directions in $I_9^-$ generates the sign-twisted representation $\mathbf{1}'\oplus\mathbf{2}$. The complete fermion space therefore decomposes as
\begin{eqnarray}
\mathcal{F}_{\Cl{8}}
&\simeq&
8\left(\mathbf{1}\oplus\mathbf{2}\right)
\oplus
8\left(\mathbf{1}'\oplus\mathbf{2}\right).
\label{eq:full-Cl8-S3-decomposition}
\end{eqnarray}

By contrast, the eight colour generators and the electromagnetic generator are fixed by the full $S_3$ family action. The fermion sector is therefore organised into three family-related sectors, while the gauge sector remains a single untriplicated copy of
\begin{eqnarray}
\SUthreeC\times\Uoneem.
\end{eqnarray}
Moreover, the gauge generators are linearly independent from the selected fermion space. Defining
\begin{eqnarray}
\mathcal{G}_{\rm gauge}
&:=&
\operatorname{span}_{\bb{C}}\{\Lambda_1,\ldots,\Lambda_8,Q\},
\end{eqnarray}
one finds
\begin{eqnarray}
\mathcal{F}_{\Cl{8}}\cap\mathcal{G}_{\rm gauge}
&=&
\{0\}.
\end{eqnarray}
Consequently,
\begin{eqnarray}
\dim_{\bb{C}}\left(
\mathcal{F}_{\Cl{8}}\oplus\mathcal{G}_{\rm gauge}
\right)
&=&
48+9
=
57.
\end{eqnarray}

This remains an unbroken-gauge construction. It establishes the algebraic family mechanism and the common electrocolour assignments, but does not yet include weak isospin or hypercharge. These structures arise in the subsequent extension to $\Cl{10}$.

\section{The $\Cl{10}$ electroweak extension and the full Standard Model gauge algebra}
\label{sec:Cl10-electroweak-extension}
The passage to $\Cl{10}$ is motivated by a structural limitation of the $\Cl{8}$ construction. In $\Cl{8}$, the colour commutator action localises appropriately on only four of the sixteen minimal left ideals, namely $I_1$, $I_{14}$, $I_9$, and $I_6$. These four ideals already provide the $S_3$-closed support required for the three-generation electrocolour construction. One cannot enlarge the fermion sectors simply by replacing the selected semi-spinors with full minimal left ideals, because the $S_3$ orbit of a full ideal is confined to a $32$-dimensional closed block and therefore cannot yield three linearly independent $16$-dimensional generation sectors. Within these compatibility requirements, $\Cl{8}$ therefore does not contain sufficient independent colour-compatible ideal support to accommodate the additional weak-doublet and weak-singlet structure. The extension to $\Cl{10}$ supplies a fifth Witt pair and the additional minimal-ideal support needed for the electroweak sector.

The extension from the electrocolour construction to the full SM gauge structure requires more than simply adjoining an additional Witt pair to $\Cl{8}$. The weak generators must commute with the existing colour algebra, be invariant under the full $S_3$ family action, and reproduce the weak-doublet/singlet structure of the SM.

A crucial constraint comes from the support of the fermion states in minimal left ideals. If $X\in\Cl{10}f$, then left multiplication by any Clifford element preserves the same ideal,
\begin{eqnarray}
AX
&\in&
\Cl{10}f.
\end{eqnarray}
This is compatible with the colour action, which acts within the selected fermion ideals. The two components of a weak doublet, however, are supported in different minimal left ideals. A weak raising or lowering operation must therefore be capable of changing the right idempotent support, which cannot be achieved by left multiplication alone. Passing from $\Cl{8}$ to $\Cl{10}$ introduces a fifth Witt pair whose right action connects the required ideals. The weak commutator action is consequently arranged to localise to right multiplication on the weak-doublet subspaces.

A further ingredient is needed to distinguish weak doublets from weak singlets. Without an additional projector, the resulting $\mathfrak{su}(2)$ action would extend to all compatible ideal sectors rather than vanishing on the singlets. The weak generators are therefore constructed using a projector $P$ onto the weak-doublet support, chosen so that
\begin{eqnarray}
[P,\Lambda_a]
&=&
0,
\qquad
a=1,\ldots,8,
\\
\psi_3(P)
&=&
\epsilon(P)
=
P.
\end{eqnarray}
Thus $P$ is colour-compatible and fixed by the full family symmetry, and its support must contain the complete $S_3$ orbit of the weak-doublet sectors.

The $\Cl{10}$ extension is therefore constrained simultaneously by colour--weak commutativity, the required left- and right-localisation of the gauge actions, the vanishing of the weak action on singlet sectors, $S_3$ invariance of the gauge generators, the SM charge assignments, and the linear independence of the three family sectors. The remainder of this section shows how these requirements are realised within a single $\Cl{10}$ algebra.

\subsection{Embedding $\Cl{8}$ into $\Cl{10}$ and the fifth Witt pair}

We realise the enlarged algebra through the graded tensor-product decomposition
\begin{eqnarray}
\Cl{10}
&\simeq&
\Cl{2}\,\widehat{\otimes}\,\Cl{8}.
\label{eq:Cl10-graded-tensor}
\end{eqnarray}
The original generators are embedded by
\begin{eqnarray}
\iota(e_i)
&=&
\mathbf{1}_2\otimes e_i,
\qquad
i=1,\ldots,8,
\end{eqnarray}
while the two additional generators may be chosen as
\begin{eqnarray}
e_9
&=&
-\ii\bar{\sigma}_1\otimes\omega_8,
\qquad
e_{10}
=
\ii\bar{\sigma}_2\otimes\omega_8.
\label{eq:e9-e10}
\end{eqnarray}
Here $\omega_8=e_1e_2\cdots e_8$ is the $\Cl{8}$ pseudoscalar and the barred Pauli matrices act on the additional $\Cl{2}$ factor. These ten generators satisfy
\begin{eqnarray}
e_Ae_B+e_Be_A
&=&
-2\delta_{AB}\mathbf{1}_{32},
\qquad
A,B=1,\ldots,10.
\end{eqnarray}

The additional generators define a fifth Witt pair,
\begin{eqnarray}
a_5
&:=&
\frac{1}{2}\left(-e_9+\ii e_{10}\right),
\qquad
a_5^\dagger
:=
\frac{1}{2}\left(e_9+\ii e_{10}\right),
\label{eq:fifth-Witt-pair}
\end{eqnarray}
which completes the five-pair Witt basis. Thus, for $i,j=1,\ldots,5$,
\begin{eqnarray}
\{a_i,a_j\}
=
\{a_i^\dagger,a_j^\dagger\}
&=&
0,
\qquad
\{a_i,a_j^\dagger\}
=
\delta_{ij}.
\label{eq:Cl10-Witt-relations}
\end{eqnarray}
In the corresponding $2\times2$ block realisation,
\begin{eqnarray}
a_5
&=&
\begin{pmatrix}
0&\ii\omega_8\\
0&0
\end{pmatrix},
\qquad
a_5^\dagger
=
\begin{pmatrix}
0&0\\
\ii\omega_8&0
\end{pmatrix}.
\label{eq:a5-block-form}
\end{eqnarray}
The fifth Witt pair therefore connects the two embedded $\Cl{8}$ blocks rather than acting independently within either one.

For $i=1,\ldots,5$, define
\begin{eqnarray}
\pi_i^{(+)}
&:=&
a_i a_i^\dagger,
\qquad
\pi_i^{(-)}
:=
a_i^\dagger a_i,
\end{eqnarray}
and
\begin{eqnarray}
f_{\varepsilon_1\varepsilon_2\varepsilon_3\varepsilon_4\varepsilon_5}
&:=&
\prod_{i=1}^{5}\pi_i^{(\varepsilon_i)},
\qquad
\varepsilon_i\in\{+,-\}.
\label{eq:Cl10-primitive-idempotents}
\end{eqnarray}
The four primitive idempotents supporting the reference fermion sector are
\begin{eqnarray}
f_{+++++}=E_{1,1},
\qquad
f_{++++-}=E_{17,17},
\qquad
f_{+++-+}=E_{9,9},
\qquad
f_{+++--}=E_{25,25}.
\label{eq:first-sector-Cl10-idempotents}
\end{eqnarray}
We denote the corresponding minimal left ideals by
\begin{eqnarray}
I_1
&:=&
\Cl{10}f_{+++++}
=
\Cl{10}E_{1,1},
\qquad
I_{17}
:=
\Cl{10}f_{++++-}
=
\Cl{10}E_{17,17},
\\
I_9
&:=&
\Cl{10}f_{+++-+}
=
\Cl{10}E_{9,9},
\qquad
I_{25}
:=
\Cl{10}f_{+++--}
=
\Cl{10}E_{25,25}.
\label{eq:first-sector-Cl10-ideals}
\end{eqnarray}

Right multiplication by the fourth and fifth Witt annihilation operators changes the corresponding idempotent occupations. At the level of minimal left ideals,
\begin{eqnarray}
I_1a_5
&=&
I_{17},
\qquad
I_9a_5
=
I_{25},
\\
I_1a_4
&=&
I_9,
\qquad
I_{17}a_4
=
I_{25}.
\label{eq:Cl10-ideal-right-maps}
\end{eqnarray}
Consequently,
\begin{eqnarray}
I_1a_4a_5
&=&
I_{25}.
\end{eqnarray}
These relations anticipate the later organisation of the weak-doublet and weak-singlet sectors: right multiplication by $a_5$ connects the two weak-isospin components, while right multiplication by $a_4$ connects the doublet and singlet supports.

The $S_3$ action is extended to $\Cl{10}$ by leaving the additional $\Cl{2}$ factor fixed. Denoting the lifted maps again by $\psi_3$ and $\epsilon$, one has
\begin{eqnarray}
\psi_3(\bar{\sigma}_a)
&=&
\bar{\sigma}_a,
\qquad
\epsilon(\bar{\sigma}_a)
=
\bar{\sigma}_a,
\qquad
a=1,2,3.
\end{eqnarray}
In the associated block-matrix realisation, the action is applied identically to the two $\Cl{8}$ blocks:
\begin{eqnarray}
\psi_3^{(10)}
&=&
\psi_3\oplus\psi_3,
\qquad
\epsilon^{(10)}
=
\epsilon\oplus\epsilon.
\label{eq:S3-Cl10-block-action}
\end{eqnarray}
We suppress the superscript when no confusion can arise.

Although the additional matrix factor is fixed, the order-three generator acts nontrivially on $e_9$ and $e_{10}$ because they contain the pseudoscalar $\omega_8$:
\begin{eqnarray}
\psi_3(e_9)
&=&
-\ii\bar{\sigma}_1\otimes\psi_3(\omega_8),
\qquad
\psi_3(e_{10})
=
\ii\bar{\sigma}_2\otimes\psi_3(\omega_8).
\label{eq:psi3-e9-e10}
\end{eqnarray}
By contrast, $\epsilon(\omega_8)=\omega_8$, and therefore
\begin{eqnarray}
\epsilon(a_5)
&=&
a_5,
\qquad
\epsilon(a_5^\dagger)
=
a_5^\dagger.
\end{eqnarray}
Since $\psi_3$ is an algebra automorphism, the transformed five-pair Witt systems again satisfy the canonical anticommutation relations. The $S_3$ family action therefore extends consistently from $\Cl{8}$ to $\Cl{10}$.

The colour generators embed as
\begin{eqnarray}
\Lambda_a
&\longmapsto&
\mathbf{1}_2\otimes\Lambda_a,
\qquad
a=1,\ldots,8,
\label{eq:Cl10-colour-embedding}
\end{eqnarray}
and remain fixed by the family action,
\begin{eqnarray}
\psi_3(\Lambda_a)
&=&
\epsilon(\Lambda_a)
=
\Lambda_a,
\qquad
a=1,\ldots,8.
\label{eq:Cl10-colour-S3-invariance}
\end{eqnarray}
The left-localised colour action established in $\Cl{8}$ is therefore inherited unchanged by the selected $\Cl{10}$ fermion subspaces.

\subsection{Selection of the electroweak fermion subspaces}

The fermion subspaces in $\Cl{10}$ are constructed as electroweak lifts of the colour-compatible semi-spinors selected in $\Cl{8}$. Their form is constrained by several requirements: the colour commutator action must remain left-localised, the weak action must localise to the right on the doublet components and vanish on the singlets, the abelian generators must reproduce the SM assignments, and the subsequent $S_3$ orbit must remain linearly independent. The resulting spaces are therefore proper subspaces of the minimal left ideals introduced above rather than arbitrary extensions obtained by adjoining the fifth Witt mode.

The natural first attempt would be to embed the $\Cl{8}$ semi-spinor $I_1^+$ directly into $\Cl{10}$ with the fifth mode unoccupied. This almost gives the required subspace, but it is not compatible with the $S_3$-invariant weak projector introduced below. In particular, the colour-singlet state $f_{+++++}$ is not annihilated by the left weak-generator term once the projector is extended over the complete $S_3$ orbit. The weak commutator would therefore fail to localise purely to the right on this state. The required subspace is obtained by replacing this single colour-singlet direction by $a_4^\dagger a_5^\dagger f_{+++++}$. This modification preserves both the dimension and the $\SU(3)_C$ representation content while allowing the weak action to localise correctly.

We therefore define the eight-dimensional subspace
\begin{eqnarray}
V_1^+
&:=&
\operatorname{span}_{\bb{C}}
\left\{
a_1^\dagger a_2^\dagger a_3^\dagger a_4^\dagger,\,
a_i^\dagger a_4^\dagger,\,
a_i^\dagger a_j^\dagger,\,
a_4^\dagger a_5^\dagger
\right\}
f_{+++++},
\label{eq:V1plus-Cl10}
\end{eqnarray}
where $i,j\in\{1,2,3\}$ with $i<j$. Its colour content is
\begin{eqnarray}
V_1^+
&\simeq&
\mathbf{1}\oplus\mathbf{3}\oplus\overline{\mathbf{3}}\oplus\mathbf{1},
\end{eqnarray}
the same as that of the corresponding $\Cl{8}$ semi-spinor.

The remaining subspaces are generated from $V_1^+$ by right multiplication with the fourth and fifth Witt annihilation operators:
\begin{eqnarray}
V_1^-
&:=&
V_1^+a_5,
\qquad
U_1^-
:=
V_1^+a_4,
\qquad
U_1^+
:=
V_1^+a_4a_5.
\label{eq:Cl10-sector-right-actions}
\end{eqnarray}
Equivalently,
\begin{eqnarray}
U_1^-
&=&
V_1^-a_4a_5^\dagger,
\qquad
U_1^+
=
V_1^-a_4,
\label{eq:Cl10-equivalent-sector-actions}
\end{eqnarray}
where these relations are understood as equalities of subspaces, so overall signs arising from anticommutation are immaterial.

The ideal-support relations established above give
\begin{eqnarray}
V_1^+
&\subset&
I_1,
\qquad
V_1^-
\subset
I_{17},
\qquad
U_1^-
\subset
I_9,
\qquad
U_1^+
\subset
I_{25}.
\label{eq:Cl10-sector-ideal-support}
\end{eqnarray}
Each subspace has complex dimension eight,
\begin{eqnarray}
\dim_{\bb{C}}V_1^+
=
\dim_{\bb{C}}V_1^-
=
\dim_{\bb{C}}U_1^-
=
\dim_{\bb{C}}U_1^+
&=&
8.
\end{eqnarray}
Since they lie in distinct minimal left ideals, their sum is direct. The reference generation sector is therefore
\begin{eqnarray}
\Gen{1}
&:=&
V_1^+\oplus V_1^-\oplus U_1^-\oplus U_1^+,
\qquad
\dim_{\bb{C}}\Gen{1}
=
32.
\label{eq:Gen1-Cl10}
\end{eqnarray}

As verified below, $V_1^+\oplus V_1^-$ carries the weak-doublet content, with $T_3=\pm\frac{1}{2}$, while $U_1^-\oplus U_1^+$ consists of weak singlets. Their detailed SM assignments follow from the colour, weak-isospin, electric-charge, and hypercharge eigenvalues derived below.

\subsection{Weak generators and the right-localised isospin action}

The weak $\mathfrak{su}(2)_L$ generators are constructed from the fifth Witt pair together with a projector onto the weak-doublet support. For the reference sector, the two doublet components are supported in the ideals generated by $f_{+++++}$ and $f_{++++-}$. Since the corresponding $\Cl{8}$ idempotent $f_{++++}$ is not fixed individually by $\psi_3$, $S_3$ invariance requires the complete closed pair $f_{++++}$ and $f_{---+}$ introduced in the previous section. After adjoining both fifth-mode occupations, this leads to
\begin{eqnarray}
P
&:=&
f_{+++++}+f_{---++}+f_{++++-}+f_{---+-}
\\
&=&
E_{1,1}+E_{14,14}+E_{17,17}+E_{30,30}.
\label{eq:P-projector}
\end{eqnarray}
In addition to $I_1$ and $I_{17}$, we therefore define
\begin{eqnarray}
I_{14}
&:=&
\Cl{10}f_{---++},
\qquad
I_{30}
:=
\Cl{10}f_{---+-}.
\end{eqnarray}
The projector $P$ is thus the minimal diagonal projector containing the reference weak-doublet support together with its complete $S_3$ orbit.

Because both fifth-mode occupations are included in $P$,
\begin{eqnarray}
[P,a_5]
&=&
[P,a_5^\dagger]
=
[P,\omega_8]
=
0.
\end{eqnarray}
Moreover, $P$ commutes with the colour algebra and is fixed by the full family symmetry:
\begin{eqnarray}
[P,\Lambda_a]
&=&
0,
\qquad
a=1,\ldots,8,
\\
\psi_3(P)
&=&
\epsilon(P)
=
P.
\label{eq:P-S3-invariance}
\end{eqnarray}

The weak generators are then defined by
\begin{eqnarray}
T_1
&:=&
\frac{1}{2}\left(-\ii a_5+\ii a_5^\dagger\right)\omega_8P
=
\frac{\ii}{2}e_9\omega_8P,
\\
T_2
&:=&
\frac{1}{2}\left(a_5+a_5^\dagger\right)\omega_8P
=
\frac{\ii}{2}e_{10}\omega_8P,
\\
T_3
&:=&
\frac{1}{2}\left(a_5^\dagger a_5-a_5a_5^\dagger\right)P
=
\frac{\ii}{2}e_9e_{10}P.
\label{eq:weak-generators-Cl10}
\end{eqnarray}
Equivalently, the ladder generators are
\begin{eqnarray}
T_+
&:=&
T_1+\ii T_2
=
\ii a_5^\dagger\omega_8P,
\qquad
T_-
:=
T_1-\ii T_2
=
-\ii a_5\omega_8P.
\label{eq:weak-ladder-generators}
\end{eqnarray}
They satisfy
\begin{eqnarray}
[T_i,T_j]
&=&
\ii\varepsilon_{ijk}T_k,
\qquad
[T_i,\Lambda_a]
=
0,
\label{eq:weak-generator-relations}
\end{eqnarray}
so the weak and colour algebras commute.

The factors of $\omega_8$ remove the dependence of $T_1$ and $T_2$ on the transformed $\Cl{8}$ pseudoscalar. Indeed,
\begin{eqnarray}
e_9\omega_8
&=&
-\ii\bar{\sigma}_1\otimes\mathbf{1}_{16},
\qquad
e_{10}\omega_8
=
\ii\bar{\sigma}_2\otimes\mathbf{1}_{16}.
\end{eqnarray}
Together with the $S_3$ invariance of $P$, this gives
\begin{eqnarray}
\psi_3(T_i)
&=&
\epsilon(T_i)
=
T_i,
\qquad
i=1,2,3.
\label{eq:weak-S3-invariance}
\end{eqnarray}
The weak gauge algebra is therefore fixed pointwise by the family action.

We now consider the commutator action on the selected fermion subspaces. For $X_+\in V_1^+$ and $X_-\in V_1^-$, their ideal support gives
\begin{eqnarray}
T_iX_+
&=&
T_iX_-
=
0,
\qquad
i=1,2,3.
\label{eq:weak-left-term-vanishes}
\end{eqnarray}
This is precisely the localisation condition that motivated the choice of $V_1^+$ in the preceding subsection. The left-multiplication term in the commutator vanishes, leaving
\begin{eqnarray}
[T_i,X_+]
&=&
-X_+T_i,
\qquad
[T_i,X_-]
=
-X_-T_i.
\label{eq:weak-right-localisation}
\end{eqnarray}
Thus weak isospin acts through right multiplication on the doublet subspaces.

Right multiplication by the fifth Witt pair changes the supporting minimal left ideal according to
\begin{eqnarray}
I_1a_5
&=&
I_{17},
\qquad
I_{14}a_5
=
I_{30},
\\
I_{17}a_5^\dagger
&=&
I_1,
\qquad
I_{30}a_5^\dagger
=
I_{14}.
\label{eq:weak-ideal-transitions}
\end{eqnarray}
Accordingly,
\begin{eqnarray}
[T_-,V_1^+]
&\subseteq&
V_1^-,
\qquad
[T_+,V_1^-]
\subseteq
V_1^+,
\\
{}[T_+,V_1^+]
&=&
0,
\qquad
[T_-,V_1^-]
=
0.
\label{eq:weak-ladder-action}
\end{eqnarray}
The diagonal generator acts as
\begin{eqnarray}
[T_3,X_+]
&=&
\frac{1}{2}X_+,
\qquad
[T_3,X_-]
=
-\frac{1}{2}X_-.
\label{eq:T3-doublet-eigenvalues}
\end{eqnarray}
Hence $V_1^+\oplus V_1^-$ carries the weak-doublet representation, with its two $T_3$ components supported in different minimal left ideals.

By contrast, the supports of $U_1^-$ and $U_1^+$ lie outside the image of $P$. Both terms in the commutator therefore vanish:
\begin{eqnarray}
[T_i,U_1^-]
&=&
[T_i,U_1^+]
=
0,
\qquad
i=1,2,3.
\label{eq:weak-singlet-action}
\end{eqnarray}
Thus $U_1^-\oplus U_1^+$ consists of weak singlets. Since the generators $T_i$ are $S_3$-invariant, the same right-localised doublet action and vanishing singlet action hold on the two $\psi_3$-generated family sectors.

\subsection{$S_3$-invariant electric charge and hypercharge}

The $S_3$-invariant electromagnetic generator $Q$ inherited from $\Cl{8}$ does not distinguish the two weak-isospin components introduced by the fifth Witt pair. Since
\begin{eqnarray}
[Q,a_5]
&=&
[Q,a_5^\dagger]
=
0,
\end{eqnarray}
any charge eigenstate $X$ satisfies
\begin{eqnarray}
[Q,X]
&=&
qX
\qquad\Longrightarrow\qquad
[Q,Xa_5]
=
qXa_5.
\end{eqnarray}
Thus corresponding states in $V_1^+$ and $V_1^-$ receive the same eigenvalue from $Q$. An additional $S_3$-invariant contribution is therefore required to produce the electric-charge splitting within each weak doublet.

The fifth-mode occupation projector is
\begin{eqnarray}
a_5^\dagger a_5
&=&
\frac{1}{2}
\left(
\mathbf{1}_{32}
+
\ii e_9e_{10}
\right).
\end{eqnarray}
Although $a_5$ and $a_5^\dagger$ are not individually fixed by $\psi_3$, their product is:
\begin{eqnarray}
\psi_3(a_5^\dagger a_5)
&=&
\epsilon(a_5^\dagger a_5)
=
a_5^\dagger a_5.
\end{eqnarray}
Together with the $S_3$-invariant weak projector $P$, this gives the invariant fifth-mode correction
\begin{eqnarray}
\Delta Q_5
&:=&
\left(
2P-\mathbf{1}_{32}
\right)
a_5^\dagger a_5.
\end{eqnarray}
Here $a_5^\dagger a_5$ detects the fifth-mode occupation, while $2P-\mathbf{1}_{32}$ supplies the required relative sign between the weak-doublet support and its complement. The electromagnetic generator in $\Cl{10}$ is therefore
\begin{eqnarray}
Q'
&:=&
Q+\Delta Q_5,
\\
&=&
Q+
\left(
2P-\mathbf{1}_{32}
\right)
a_5^\dagger a_5.
\label{eq:Qprime-Cl10}
\end{eqnarray}
Since both terms are fixed by the family action,
\begin{eqnarray}
\psi_3(Q')
=
\epsilon(Q')
=
Q'.
\label{eq:Qprime-S3-invariance}
\end{eqnarray}

The hypercharge generator is then defined by the SM relation
\begin{eqnarray}
Y
&:=&
2\left(Q'-T_3\right),
\qquad
Q'
=
T_3+\frac{1}{2}Y.
\label{eq:hypercharge-Cl10}
\end{eqnarray}
Direct evaluation gives
\begin{eqnarray}
\psi_3(Y)
&=&
\epsilon(Y)
=
Y,
\\
{}[Q',\Lambda_a]
&=&
[Y,\Lambda_a]
=
0,
\qquad
a=1,\ldots,8,
\\
{}[Y,T_i]
&=&
0,
\qquad
i=1,2,3.
\label{eq:hypercharge-commutation}
\end{eqnarray}
Moreover,
\begin{eqnarray}
[Q',T_\pm]
&=&
\pm T_\pm,
\end{eqnarray}
as required for the electrically charged weak ladder generators. Hence $Y$ generates a family-invariant $\mathfrak{u}(1)_Y$ commuting with both $\mathfrak{su}(3)_C$ and $\mathfrak{su}(2)_L$.

The joint eigenvalues of $T_3$, $Q'$, and $Y$, together with the colour representations, reproduce the fermion gauge quantum numbers of one SM generation on the reference sector $\Gen{1}$. The explicit correspondence is summarised in Table~\ref{tab:Cl10-SM-states}.

\begin{table}[t]
\centering
\small
\renewcommand{\arraystretch}{1.12}
\begin{tabular}{c c c c c c}
\hline
Sector & Algebraic state & Interpretation & $\SUthreeC$ & $T_3$ & $(Q',Y)$ \\
\hline
$V_1^+$ & $a_1^\dagger a_2^\dagger a_3^\dagger a_4^\dagger f_{+++++}$ & $e_L^+$ & $\mathbf{1}$ & $+\frac{1}{2}$ & $(+1,+1)$ \\
$V_1^+$ & $a_i^\dagger a_j^\dagger f_{+++++}$ & $\overline{d}_L$ & $\overline{\mathbf{3}}$ & $+\frac{1}{2}$ & $(+\frac{1}{3},-\frac{1}{3})$ \\
$V_1^+$ & $a_i^\dagger a_4^\dagger f_{+++++}$ & $u_L$ & $\mathbf{3}$ & $+\frac{1}{2}$ & $(+\frac{2}{3},+\frac{1}{3})$ \\
$V_1^+$ & $a_4^\dagger a_5^\dagger f_{+++++}$ & $\nu_L$ & $\mathbf{1}$ & $+\frac{1}{2}$ & $(0,-1)$ \\
\hline
$V_1^-$ & $a_1^\dagger a_2^\dagger a_3^\dagger a_4^\dagger a_5f_{++++-}$ & $\overline{\nu}_L$ & $\mathbf{1}$ & $-\frac{1}{2}$ & $(0,+1)$ \\
$V_1^-$ & $a_i^\dagger a_j^\dagger a_5f_{++++-}$ & $\overline{u}_L$ & $\overline{\mathbf{3}}$ & $-\frac{1}{2}$ & $(-\frac{2}{3},-\frac{1}{3})$ \\
$V_1^-$ & $a_i^\dagger a_4^\dagger a_5f_{++++-}$ & $d_L$ & $\mathbf{3}$ & $-\frac{1}{2}$ & $(-\frac{1}{3},+\frac{1}{3})$ \\
$V_1^-$ & $a_4^\dagger f_{++++-}$ & $e_L^-$ & $\mathbf{1}$ & $-\frac{1}{2}$ & $(-1,-1)$ \\
\hline
$U_1^+$ & $a_1^\dagger a_2^\dagger a_3^\dagger a_5f_{+++--}$ & $e_R^+$ & $\mathbf{1}$ & $0$ & $(+1,+2)$ \\
$U_1^+$ & $a_i^\dagger a_j^\dagger a_4a_5f_{+++--}$ & $\overline{d}_R$ & $\overline{\mathbf{3}}$ & $0$ & $(+\frac{1}{3},+\frac{2}{3})$ \\
$U_1^+$ & $a_i^\dagger a_5f_{+++--}$ & $u_R$ & $\mathbf{3}$ & $0$ & $(+\frac{2}{3},+\frac{4}{3})$ \\
$U_1^+$ & $f_{+++--}$ & $\nu_R$ & $\mathbf{1}$ & $0$ & $(0,0)$ \\
\hline
$U_1^-$ & $a_1^\dagger a_2^\dagger a_3^\dagger f_{+++-+}$ & $\overline{\nu}_R$ & $\mathbf{1}$ & $0$ & $(0,0)$ \\
$U_1^-$ & $a_i^\dagger a_j^\dagger a_4f_{+++-+}$ & $\overline{u}_R$ & $\overline{\mathbf{3}}$ & $0$ & $(-\frac{2}{3},-\frac{4}{3})$ \\
$U_1^-$ & $a_i^\dagger f_{+++-+}$ & $d_R$ & $\mathbf{3}$ & $0$ & $(-\frac{1}{3},-\frac{2}{3})$ \\
$U_1^-$ & $a_5^\dagger f_{+++-+}$ & $e_R^-$ & $\mathbf{1}$ & $0$ & $(-1,-2)$ \\
\hline
\end{tabular}
\caption{Algebraic representatives and gauge quantum numbers in the reference fermion sector $\Gen{1}$. Here $i,j\in\{1,2,3\}$ with $i<j$, and the final column lists $(Q',Y)$.}
\label{tab:Cl10-SM-states}
\end{table}

The labels $L$ and $R$ in Table~\ref{tab:Cl10-SM-states} refer to the corresponding chiral SM gauge representations. Each generation sector contains two algebraically distinct neutral singlets, interpreted as a right-handed neutrino and its antiparticle. Both are singlets under the full SM gauge algebra, so the right-handed neutrino is sterile with respect to the SM gauge interactions. Their distinction as $\nu_R$ and $\overline{\nu}_R$ relies on the particle--antiparticle organisation of the algebraic state space rather than on the gauge action alone. The construction does not intrinsically identify $\Gen{1}$ with the observed electron generation rather than the muon or tau generation.

\subsection{Three generations and an untriplicated gauge sector}

The remaining two fermion sectors are generated from the $32$-dimensional reference sector by the order-three family action:
\begin{eqnarray}
\Gen{2}
:=
\psi_3(\Gen{1}),
\qquad
\Gen{3}
:=
\psi_3^2(\Gen{1}).
\label{eq:Cl10-three-generations}
\end{eqnarray}
In the explicit matrix realisation,
\begin{eqnarray}
\dim_{\bb{C}}
\left(
\Gen{1}+\Gen{2}+\Gen{3}
\right)
&=&
96
\\
&=&
\dim_{\bb{C}}\Gen{1}
+
\dim_{\bb{C}}\Gen{2}
+
\dim_{\bb{C}}\Gen{3}.
\label{eq:Cl10-generation-rank}
\end{eqnarray}
The three sectors are therefore linearly independent, and the fermion space is
\begin{eqnarray}
\mathcal{F}_{\Cl{10}}
&:=&
\Gen{1}\oplus\Gen{2}\oplus\Gen{3}.
\end{eqnarray}

Every generator of
\begin{eqnarray}
\mathfrak{g}_{\rm SM}
&:=&
\mathfrak{su}(3)_C
\oplus
\mathfrak{su}(2)_L
\oplus
\mathfrak{u}(1)_Y
\end{eqnarray}
is fixed by the full $S_3$ family action. Corresponding states in $\Gen{1}$, $\Gen{2}$, and $\Gen{3}$ therefore carry identical SM gauge quantum numbers. On the complete fermion space, the common adjoint gauge action is left-localised for colour, right-localised for weak isospin on the doublet subspaces, and diagonal for electric charge and hypercharge.

The twelve gauge generators are also linearly independent from the $96$-dimensional fermion space. Defining
\begin{eqnarray}
\mathcal{G}_{\rm SM}
&:=&
\operatorname{span}_{\bb{C}}
\left\{
\Lambda_1,\ldots,\Lambda_8,
T_1,T_2,T_3,Y
\right\},
\end{eqnarray}
one finds
\begin{eqnarray}
\mathcal{F}_{\Cl{10}}
\cap
\mathcal{G}_{\rm SM}
&=&
\{0\},
\end{eqnarray}
and hence
\begin{eqnarray}
\dim_{\bb{C}}
\left(
\mathcal{F}_{\Cl{10}}
\oplus
\mathcal{G}_{\rm SM}
\right)
&=&
96+12
=
108.
\label{eq:Cl10-fermion-gauge-rank}
\end{eqnarray}
Thus the $S_3$ family symmetry relates three linearly independent fermion sectors while leaving a single untriplicated $\SUthreeC\times\SUtwoL\times\UoneY$ gauge sector.

\section{Comparison with other three-generation mechanisms}
\label{sec:comparative-analysis}


Having presented the sedenion-motivated $\Cl{8}$ and $\Cl{10}$ framework, we now place its generation mechanism in a broader context by comparing it with representative approaches in which fermion-family replication is related to division algebras, Clifford algebras, $\Spin(8)$ triality, exceptional Jordan algebras, dimensional reduction, or an $S_3$ family symmetry. The purpose is not to survey algebraic particle physics as a whole, but to distinguish the different algebraic mechanisms that give rise to a threefold organisation.

Three questions guide the comparison. First, what algebraic feature produces the threefold structure, and does an explicit symmetry relate the proposed generations? Second, what fermion and gauge content is obtained, and do distinct fermion sectors transform equivalently under a common gauge algebra? Third, which ingredients are assumed rather than explained by the construction, and what limitations remain in interpreting its threefold structure as the three observed SM generations? These criteria provide a common basis for comparison without assuming that the different approaches have the same aims or make the same physical claims.

\subsection{Division-algebra and Clifford-algebra constructions}
\label{subsec:division-clifford-comparison}

Closely related approaches use division algebras to generate associative algebras of multiplication operators whose representation content can then be compared with that of the SM. As reviewed in Section~\ref{sec:division-algebraic-ladder-operators}, these methods provide economical descriptions of the internal quantum numbers of one fermion generation. Several extensions have also produced threefold patterns, although the origin and physical interpretation of the number three differ.

An early proposal for family replication was given by Dixon, as reviewed in Section~\ref{sec:dixon-algebra}. Beginning with a one-family hyperspinor space based on $\bb{T}^2$, Dixon proposed the enlarged structure
\begin{eqnarray}
\bb{T}^6&=&\bb{C}^{1}\otimes\bb{H}^{2}\otimes\bb{O}^{3}.
\end{eqnarray}
The number three is encoded in the chosen multiplicities of the division algebra factors and is motivated by their distinguished mathematical properties, rather than generated by a symmetry relating three fermion sectors. Moreover, the full multiplication algebra associated with $\bb{T}^6$ is not itself a Clifford algebra. The construction therefore provides a structured proposal for family replication, but not a Clifford-ideal realisation of three generations related by an explicit family transformation.

Furey subsequently obtained the electrocolour representation content of three generations from the complex octonions and their associated left-action algebra $\Cl{6}$ \cite{Furey2014,Furey2018ThreeGenerations}. Decomposing the full $64$-complex-dimensional algebra under a selected colour action produces the $\SUthreeC$ representations of three generations of quarks and leptons; the later extension includes electric charge and hence the full $\SUthreeC\times\Uoneem$ content. Here the number three arises from the multiplicities with which the required representations occur in the decomposition of $\Cl{6}$. The construction does not introduce a family transformation relating the three sets, and it does not include the chiral weak interaction. Its scope is therefore the electrocolour representation content of three generations rather than three complete generations transforming under the full SM gauge group.

A separate construction based on five division-algebraic ladder pairs selects the globally correct SM gauge group from a larger symmetry resembling that of $\SU(5)$ \cite{Furey2018a}. Requiring transformations to preserve the distinction between the underlying algebraic actions removes those associated with proton decay and leaves $\bigl[\SUthreeC\times\SUtwoL\times\UoneY\bigr]/\bb{Z}_6$, with a possible additional $\U(1)_X$. This result concerns the origin of the SM gauge symmetry for one generation and is not combined, within the same construction, with the three-generation decomposition of $\Cl{6}$.

More recent work by Furey organises particle representations within the $256$-real-dimensional Jordan algebra $\mathcal{H}_{16}(\bb{C})$ \cite{Furey2025Superalgebra}. A distinguished decomposition separates the algebra into one sector associated with a derivative operator and three disjoint sectors associated with fermion generations. The number three therefore arises from this decomposition rather than from an explicit family symmetry relating the fermion sectors. The construction contains representations corresponding to the SM gauge bosons and to most of the states of three fermion generations, but the representations involving the top quark are absent, and the symmetry preserving the decomposition contains four additional $\mathfrak{u}(1)$ factors. It consequently provides a substantial but not yet complete algebraic organisation of three-generation particle content.

Furey has also investigated gauge-symmetry selection from the nested inclusions
\begin{eqnarray}
\bb{R}\subset\bb{C}\subset\bb{H}\subset\bb{O}\subset\bb{V},
\end{eqnarray}
where $\bb{V}$ may be chosen as the sedenions and $\operatorname{End}_{\bb{R}}(\bb{V})\simeq C\ell(0,8)$ \cite{Furey2026Nested}. Conditions derived from the nested subalgebras select both the full SM gauge algebra and its unbroken electrocolour subalgebra. The simultaneous appearance of a $16$-dimensional algebra and an associated Clifford algebra makes this construction structurally relevant here, although its purpose is gauge-symmetry selection rather than family replication.

Quinta assigns a different physical role to the real Clifford algebra $C\ell(8,0)$, whose complexification is isomorphic to $\Cl{8}$ \cite{quinta2025spacetime}. Starting from the ordinary four-dimensional Dirac theory, he embeds its Clifford algebra into $C\ell(8,0)$ and interprets the larger algebra as describing an eight-dimensional spacetime. Its internal transformations produce the SM gauge group together with $\U(1)_{B-L}$, while the fermion representations occur in four Dirac-spinor families. The first three transform together under a continuous $\U(3)_F$ family symmetry and may mix with one another, whereas the fourth is a singlet under the corresponding $\SU(3)_F$ action and does not mix with them. The current formulation is therefore a $3+1$ family model rather than a construction of exactly three generations. Its principal strength is that the fermion representations, full SM gauge structure, family symmetry, and elements of the mass sector are treated within a common Clifford algebra. The central unresolved issue for the generation problem is the physical interpretation of the additional family.

\subsection{Triality-based constructions}
\label{subsec:triality-comparison}

Triality is attractive in the context of the generation problem because it provides an intrinsic occurrence of the number three. The outer automorphism group of $\Spin(8)$ permutes its vector and two inequivalent chiral-spinor representations, conventionally denoted by $\mathbf{8}_v$, $\mathbf{8}_s$, and $\mathbf{8}_c$. These representations have the same dimension but are not equivalent as $\Spin(8)$ modules. Additional structure is therefore required before they can be interpreted as three fermion generations carrying identical SM quantum numbers.

An early proposal along these lines was developed by Silagadze \cite{silagadze1994so}. His construction places $\Spin(8)$ triality within an $E_6$ setting containing three equivalent $\Spin(10)$ subgroups permuted by the triality action. Each $\Spin(10)$ subgroup can organise one quark--lepton family, but triality provides no preferred choice among the three. Silagadze therefore introduced three copies of the fundamental $\mathbf{27}$ representation of $E_6$, with a different $\Spin(10)$ subgroup playing the family-forming role in each copy. The number three is motivated by the triality-related embeddings, but three SM generations are not obtained within a single irreducible representation.

A more recent implementation was developed by Furey and Hughes using the simultaneous triality structures associated with $\bb{C}$, $\bb{H}$, and $\bb{O}$ \cite{furey2025three}. The full SM internal gauge algebra acts on a triality triple $(\Psi_+,\Psi_-,V)$ with entries in $\bb{C}\otimes\bb{H}\otimes\bb{O}$. The two spinor components, $\Psi_+$ and $\Psi_-$, directly provide the irreducible representations of two SM generations, including sterile neutrinos. The vector component $V$, however, initially decomposes into scalar rather than fermion representations.

The third generation is obtained through a Cartan factorisation, meaning a triality-compatible relation that expresses vector-representation data in terms of spinor and conjugate-spinor data. This allows the representation content associated with $V$ to be reorganised as that of a third fermion generation, with the same SM gauge algebra acting on all three components. The generations nevertheless enter asymmetrically: two occur directly in the chiral-spinor components, while the third is recovered from the vector component through the additional factorisation. The role of this factorisation and of the accompanying scalar representations in a dynamical theory remains open.

Lisi uses triality differently, by adjoining an order-three transformation $t$ to the discrete $C$, $P$, and $T$ transformations of Dirac fermions \cite{lisi2024c}. The resulting finite group, denoted the $CPTt$ group, cycles explicitly between three copies of a given Dirac fermion. The construction therefore begins with known fermion types and organises their generation, particle--antiparticle, spin, and helicity states under an enlarged discrete symmetry. It does not by itself derive the internal SM gauge group or combine the separately treated fermion types into a complete three-generation particle model.

Triality also contributes to the family structure of Quinta’s model, discussed in the preceding subsection, in which the four-dimensional Dirac theory is embedded into a Clifford algebra description of eight-dimensional spacetime \cite{quinta2025spacetime}. Three of its four Dirac-spinor families are associated with the vector and two chiral-spinor positions of $\Spin(8)$ and transform together under the continuous family symmetry $\U(3)_F$. The fourth family is a singlet under the corresponding $\SU(3)_F$ action. Triality therefore distinguishes a threefold subsystem, but it does not determine the total number of families, which is four.

These constructions show that triality does not lead to a unique generation mechanism. Depending on the model, it permutes unified-subgroup embeddings, relates vector and spinor positions, or acts on three pre-existing copies of a fermion. In each case, further structure is required to obtain equivalent SM family representations and to specify how the triality operation is realised physically.

The $S_3$ action reviewed in this article should not presently be identified with the standard triality action of $\Spin(8)$. Let
\begin{eqnarray}
V&=&\operatorname{span}_{\bb{C}}{e_1,\ldots,e_8}
\end{eqnarray}
denote the canonical one-vector generating space of $\Cl{8}$. Standard triality acts as an outer automorphism of the associated bivector algebra
\begin{eqnarray}
\mathfrak{so}(8,\bb{C})&\simeq&\Lambda^2V
\end{eqnarray}
and permutes the vector and two chiral-spinor representations. By contrast, the sedenion-induced generator $\psi_3$ is an automorphism of the full Clifford algebra but does not preserve $V$: it maps the original Clifford generators to combinations containing multivectors of different grades. Consequently, it does not preserve the canonical bivector algebra $\Lambda^2V$ or induce the standard outer triality automorphism on this copy of $\mathfrak{so}(8,\bb{C})$. Its family interpretation instead arises from its action on selected fermion subspaces inside $\Cl{8}$. Whether this action admits a reformulation in triality-adapted variables, or whether the two $S_3$ structures arise from a larger common construction, remains an open question.

\subsection{Exceptional Jordan algebra approaches}
\label{subsec:jordan-comparison}

The exceptional Jordan algebra $J_3(\bb{O})$, consisting of $3\times3$ Hermitian matrices over the octonions, provides a direct algebraic setting in which a threefold structure is present from the outset. Its elements contain three real diagonal entries and three off-diagonal octonion entries, so
\begin{eqnarray}
\dim_{\bb{R}}J_3(\bb{O})=3+3\dim_{\bb{R}}\bb{O}=27.
\end{eqnarray}
Even after complexification, the algebra has complex dimension $27$, whereas three complete SM generations require $45$ complex internal fermion states, or $48$ when right-handed neutrinos are included. The algebra therefore cannot by itself serve as the direct fermion state space of three complete generations. Approaches based on $J_3(\bb{O})$ must instead use it as an organising or observable algebra, introduce additional copies or modules, or pass to a larger exceptional or Jordan--Clifford structure.

The automorphism group of $J_3(\bb{O})$ is $F_4$, and the subgroup that fixes the three diagonal primitive idempotents is $\Spin(8)$. The three off-diagonal octonion sectors transform as the vector and two chiral-spinor representations $\mathbf{8}_v$, $\mathbf{8}_s$, and $\mathbf{8}_c$, while permutations of the diagonal positions induce the outer $S_3$ triality action \cite{todorov2018octonions,boyle2020standard2}. The rank-three structure therefore provides a natural candidate for a generation mechanism, although the three triality representations are inequivalent as $\Spin(8)$ modules.

Dubois-Violette initially proposed identifying the three off-diagonal octonion entries with three fermion generations, using the decomposition $\bb{O}=\bb{C}\oplus\bb{C}^3$ to distinguish leptons from quarks within each generation \cite{dubois2016exceptional1}. Simultaneously accommodating the up-type and down-type sectors required two copies of $J_3(\bb{O})$, however, and introduced additional diagonal fermionic states without yet giving a satisfactory electroweak structure. Dubois-Violette and Todorov subsequently associated the generations with three canonical and overlapping subalgebras $J^{(i)}\simeq J_2(\bb{O})$, each formed from a different pair of rows and columns of $J_3(\bb{O})$ \cite{dubois2019exceptional2}. Each $J^{(i)}$ is interpreted as the internal observable algebra associated with one generation.

A particular strength of this approach is that the SM gauge group is selected through geometrically distinguished stabiliser subgroups. Choosing a complex subalgebra $\bb{C}\subset\bb{O}$ and fixing a diagonal primitive idempotent determine two such subgroups whose connected intersection is
\begin{eqnarray}
\mathrm{S}\bigl(\U(2)\times\U(3)\bigr)
&\simeq&
\frac{\SUthreeC\times\SUtwoL\times\UoneY}{\bb{Z}_6},
\label{eq:exceptional-jordan-sm-intersection}
\end{eqnarray}
the globally correct SM gauge group \cite{todorov2018deducing,todorov2018octonions}. Baez and Schwahn have recently reformulated this result in terms of the nested Jordan subalgebras $J_2(\bb{C})\subset J_3(\bb{C})\subset J_3(\bb{O})$ \cite{baez2026standard}. This gives an intrinsic characterisation of the SM gauge group, but is principally a gauge-group result rather than a construction of three fermion generations. In the three-subalgebra interpretation, the $J^{(i)}$ are associated with three distinct SM gauge-group embeddings. These embeddings share the same colour subgroup but differ in their electroweak factors, rather than defining one common SM gauge embedding for all three proposed generations.

Boyle developed a related proposal based on the complexified exceptional Jordan algebra $J_3^{\bb{C}}(\bb{O})=\bb{C}\otimes J_3(\bb{O})$ and its $E_6$ symmetry \cite{boyle2020standard2}. Distinguished subgroup intersections select a minimal left--right extension of the SM gauge group. Under $\Spin(10)\subset E_6$, the fundamental representation decomposes as
\begin{eqnarray}
\mathbf{27}&=&\mathbf{1}\oplus\mathbf{10}\oplus\mathbf{16},
\label{eq:exceptional-jordan-e6-decomposition}
\end{eqnarray}
where the $\mathbf{16}$ carries the fermion representation of one SM generation, including a right-handed neutrino. The exceptional structure also contains three copies of a complex-octonion space related by $\Spin(8)$ triality, suggesting a possible origin for three generations. It does not yet assemble these spaces into an explicit direct sum of three equivalent SM-generation representations. Recent work by Baez, Bokor, and Boyle gives a more intrinsic formulation of the one-generation space $(\bb{C}\otimes\bb{O})^2$ within exceptional Jordan geometry \cite{baez2026jordanpair}. This clarifies the mathematical status of the one-generation representation but does not by itself construct three generations.

Perelman instead considers the Jordan--Clifford algebra $J_3(\bb{C}\otimes\bb{O})\otimes\Cl{4}$ and argues that it contains the spinorial degrees of freedom of three generations and their antiparticles, together with associated scalar degrees of freedom \cite{perelman2021jordan}. The internal representations are organised under $\SUthreeC\times\SUtwoL\times\SU(2)_R\times\U(1)$, while the $\Cl{4}$ factor supplies the four-dimensional spinor structure. This gives a relatively explicit accounting of the proposed matter content, although the gauge-boson sector is treated separately and the equivalence of the generations is not expressed through an intrinsic family-permutation symmetry.

Another line of work uses eigenvalues and other characteristic data of complexified exceptional Jordan matrices to investigate charged-fermion mass ratios and mixing \cite{bhatt2022majorana,singh2025fermionmass,teli2026masshierarchies,teli2026mixing}. These studies seek to connect algebraic invariants with flavour observables and therefore go beyond representation counting. The more recent analyses introduce fitted parameters and effective assumptions in passing from mass hierarchies to mixing angles. They are consequently best viewed as phenomenological extensions of the exceptional Jordan framework rather than parameter-free consequences of $J_3(\bb{O})$ alone.

Across these approaches, the number three originates primarily in the rank-three geometry of $J_3(\bb{O})$ and its relation to $\Spin(8)$ triality. This structure provides distinguished exceptional symmetries and powerful stabiliser descriptions of the SM or left--right gauge groups. The remaining challenge is to turn the three inequivalent triality sectors, or the three overlapping Jordan subalgebras, into three equivalent SM fermion generations with a clearly specified common gauge action. Although exceptional Jordan and sedenion-motivated constructions both involve $\Cl{8}$-related structures and an $S_3$ action, the Jordan-algebra $S_3$ is the standard triality permutation discussed in Section~\ref{subsec:triality-comparison}, and should not be identified with the sedenion-induced family action.

\subsection{Octonion dimensional reduction and the early sedenion proposal}
\label{subsec:dimensional-reduction-comparison}

The dimensional-reduction mechanism of Manogue and Dray, reviewed in Section~\ref{sec:manogue-dray-dimensional-reduction}, was an important precursor to the present framework, although it was developed for a different purpose \cite{manogue1999dimensional}. The choice of a preferred imaginary octonion unit selects a complex subalgebra $\bb{C}\subset\bb{O}$ and reduces the ten-dimensional Lorentz structure to a four-dimensional $SL(2,\bb{C})$ subgroup. The three quaternion subalgebras containing this preferred complex subalgebra are then interpreted as three leptonic generations. The number three therefore arises from the nested structure $\bb{C}\subset\bb{H}\subset\bb{O}$ associated with the dimensional reduction, rather than from $\Spin(8)$ triality or an explicit family-permutation symmetry.

The construction is primarily a proposal for relating spacetime dimensional reduction to a threefold organisation of leptons. It does not attempt to accommodate three complete quark--lepton generations or derive their transformation properties under a common internal SM gauge group. Moreover, the three quaternionic sectors are selected by their shared complex subalgebra, but are not related by an explicitly identified family symmetry.

This threefold arrangement directly motivated the earliest sedenion proposal, in which three octonion subalgebras of $\bb{S}$ were considered as possible carriers of three complete fermion generations. Subsequent work showed that this direct subalgebraic identification did not naturally provide independent family spaces with a common gauge action. The refined formulation therefore shifted from the overlapping octonion subalgebras themselves to the intrinsic $S_3$ automorphism structure of the sedenions and its action within $\Cl{8}$ and $\Cl{10}$.

\subsection{Relation to conventional $S_3$ family-symmetry models}
\label{subsec:s3-family-comparison}

The permutation group $S_3$ has long been used as a family or flavour symmetry because it is the smallest nonabelian group acting naturally on three objects. Representative treatments include those of González Canales, Mondragón, and Mondragón and of Kaneko et al.\ \cite{gonzalez2013s3,kaneko2007flavor}. In such models, the three copies of each fermion multiplet are assigned to representations of $S_3$. The natural three-dimensional permutation representation decomposes as $\mathbf{3}_{\rm perm}\simeq\mathbf{1}\oplus\mathbf{2}$, so one family is frequently assigned to a singlet and the remaining two to a doublet. The left-handed fermions, right-handed fermions, and scalar fields may be assigned independently, subject to the requirement that the relevant interactions are $S_3$ invariant. The imposed symmetry then constrains the possible Yukawa matrices, scalar potential, and fermion mass textures.

In the simplest exact permutation-symmetric limit, the doublet components are degenerate. Realistic mass hierarchies and mixing therefore require spontaneous, soft, or sequential breaking of $S_3$, often together with an enlarged scalar sector. González Canales et al.\ have studied fermion mixing in a three-Higgs-doublet model, Cogollo and Silva considered an $S_3$-symmetric two-Higgs-doublet model, and Babu, Wu, and Xu developed a more recent three-Higgs-doublet treatment of fermion masses, neutrino mixing, and flavour-changing processes \cite{canales2013fermion,cogollo2016two,babu2024fermion}. In these constructions, the three fermion families and their common SM gauge transformation properties are introduced at the outset. The role of $S_3$ is to organise the families and constrain their flavour dynamics, rather than to account for their existence.

The logical order is different in the framework reviewed here. The $S_3$ symmetry is inherited from the automorphism structure of the sedenions and realised on fermion subspaces inside $\Cl{8}$ and $\Cl{10}$ \cite{Gresnigt2023,Gresnigt2026PLB}. Its order-three element cyclically relates the three sectors $\Gen{1}$, $\Gen{2}$, and $\Gen{3}$. At the level of the generation labels, this orbit carries the same permutation representation $\mathbf{1}\oplus\mathbf{2}$ used in conventional $S_3$ models. At the level of the individual Clifford spinors, however, the representation content is fixed more specifically: the two semi-spinor sectors carry $\mathbf{1}\oplus\mathbf{2}$ and $\mathbf{1}'\oplus\mathbf{2}$, as shown in Eq.~\eqref{eq:full-Cl8-S3-decomposition}. These assignments follow from the embedded algebraic action rather than being chosen independently for the different fermion and scalar fields.

The relation between the family and gauge symmetries also enters differently. In conventional flavour models, the SM gauge action is already common to the three families, allowing the imposed $S_3$ symmetry to commute with it. In the present construction, the relevant gauge generators are selected to be invariant under the same algebraic $S_3$ action that relates the fermion sectors. The distinction therefore concerns not only the origin of the group $S_3$, but also the stage at which the family structure and its compatibility with the gauge action enter the construction.

The two uses of $S_3$ are nevertheless complementary. Conventional models provide well-developed methods for breaking the family symmetry and constructing phenomenologically viable masses and mixing, whereas the present framework seeks an algebraic origin for the family action itself. It does not yet derive the observed fermion mass hierarchy, mixing matrices, or CP-violating phases. Conventional $S_3$ flavour models may therefore provide useful guidance for developing this dynamical sector, provided that any extension remains compatible with the family representations and gauge invariance fixed by the Clifford algebra construction.

\subsection{Comparative synthesis}
\label{subsec:comparative-synthesis}

The approaches reviewed above employ several mathematically distinct notions of threefold structure. The number three may arise from representation multiplicities, $\Spin(8)$ triality, the rank-three structure of an exceptional Jordan algebra, nested division subalgebras, or a permutation symmetry acting on three pre-existing fermion copies. These occurrences should not be regarded as instances of a single generation mechanism. The relevant questions are how the threefold structure arises, whether a symmetry relates the proposed family sectors, and what fermion and gauge content is actually obtained. Table~\ref{tab:three-generation-mechanisms-comparison} summarises these points for the principal classes of construction considered here.

\begin{table}[p]
\centering
\footnotesize
\renewcommand{\arraystretch}{1.15}
\setlength{\arrayrulewidth}{0.3pt}
\begin{tabular}{@{}p{0.16\textwidth}|p{0.22\textwidth}|p{0.23\textwidth}|p{0.27\textwidth}@{}}
\toprule
\textbf{Approach} &
\textbf{Source of the threefold structure} &
\textbf{Relation among the proposed families} &
\textbf{Fermion and gauge scope} \\
\midrule

$\Cl{6}$ decomposition \cite{Furey2014,Furey2018ThreeGenerations}
&
Multiplicity of the relevant representations in the decomposition of the full Clifford algebra.
&
No explicit family transformation relates the three generation-like sets.
&
Reproduces the $\SUthreeC\times\Uoneem$ representation content of three generations; the chiral weak sector is not included.
\\[0.8ex]

Triality-based constructions \cite{silagadze1994so,lisi2024c,furey2025three,quinta2025spacetime}
&
The vector and two chiral-spinor positions of $\Spin(8)$, or related triality triples.
&
Triality permutes three positions, although these may be inequivalent representations or play different roles in the construction.
&
The outcome depends on the model and includes $E_6$ embeddings, a common SM action on a triality triple, transformations among assumed Dirac fields, and a $3+1$ family structure.
\\[0.8ex]

Exceptional Jordan algebra approaches \cite{dubois2019exceptional2,boyle2020standard2}
&
The rank-three Jordan structure, its three off-diagonal octonion entries, three overlapping $J_2(\bb{O})$ subalgebras, or triality-related spaces.
&
Frame permutations and triality relate the threefold structures, but do not generally produce three equivalent fermion modules.
&
Distinguished stabilisers yield the SM or left--right gauge groups. The $27$-dimensional algebra serves as an organising or observable structure rather than a direct state space for three complete generations.
\\[0.8ex]

Octonion dimensional reduction \cite{manogue1999dimensional}
&
Three quaternion subalgebras containing a preferred complex subalgebra of $\bb{O}$.
&
The sectors share the selected complex subalgebra but are not related by an explicitly identified family-permutation symmetry.
&
Connects four-dimensional Lorentz reduction with three leptonic sectors; complete quark--lepton generations and their internal SM gauge action are not constructed.
\\[0.8ex]

Conventional $S_3$ models \cite{gonzalez2013s3,babu2024fermion}
&
Three fermion families are introduced and usually assigned according to $\mathbf{3}_{\rm perm}\simeq\mathbf{1}\oplus\mathbf{2}$.
&
An imposed $S_3$ symmetry acts on the three pre-existing families.
&
The common SM gauge action is assumed, while $S_3$ constrains Yukawa interactions, scalar potentials, masses, and mixing.
\\[0.8ex]

Present framework \cite{Gresnigt2023,Gresnigt2026PLB}
&
The intrinsic $S_3$ automorphism structure of the sedenions, realised inside $\Cl{8}$ and $\Cl{10}$.
&
The order-three element cyclically relates three distinct fermion sectors.
&
The $\Cl{8}$ construction gives their common electrocolour action, while the $\Cl{10}$ extension gives a common SM gauge algebra whose generators are invariant under $S_3$.
\\
\bottomrule
\end{tabular}
\caption{Comparison of the origins, family relations, and fermion and gauge scope of representative three-generation mechanisms. The table summarises the aspects relevant to the present comparison rather than attempting an exhaustive classification of the corresponding models.}
\label{tab:three-generation-mechanisms-comparison}
\end{table}

The table separates three claims that can otherwise be easily conflated: that an algebraic construction contains three occurrences of suitable representation content, that an explicit symmetry relates three proposed family sectors, and that the same SM gauge algebra acts on all three. Different approaches establish different combinations of these statements and often address different physical questions. Representation multiplicities and rank-three geometries provide a natural occurrence of the number three, while triality supplies an explicit permutation whose three positions need not have the same algebraic status. Conventional $S_3$ models instead begin with three equivalent SM families and use the permutation symmetry to constrain their flavour dynamics.

Within this comparison, the distinguishing feature of the sedenion-motivated construction is the simultaneous realisation of the fermion sectors, gauge generators, and family action within a single associative Clifford algebra. The comparison does not imply that the other approaches have the same objectives, nor does it establish the dynamical or phenomenological completeness of the present framework. Its purpose is to identify precisely which part of the generation problem each construction addresses. The remaining limitations and open questions of the sedenion-motivated framework are considered in Section~\ref{sec:conclusions-outlook}.

\section{Conclusions and outlook}
\label{sec:conclusions-outlook}

The framework reviewed here brings together the sedenions, the complex Clifford algebras $\Cl{8}$ and $\Cl{10}$, and an intrinsic $S_3$ family symmetry. The sedenions play two related roles. Their complexified left-multiplication maps generate an associative endomorphism algebra isomorphic to $\Cl{8}$, while their automorphism structure contains the $S_3$ factor used to construct the family action. The fermion subspaces, gauge generators, and family action can therefore be represented within a common associative algebra, even though the original motivation comes from a nonassociative Cayley--Dickson algebra.

In the refined $\Cl{8}$ construction, the order-three generator $\psi_3$ produces three linearly independent fermion sectors from a single reference sector. The colour and electromagnetic generators are invariant under the family action, so the three sectors carry equivalent quantum numbers under one common $\SUthreeC\times\Uoneem$ gauge algebra. The extension to $\Cl{10}$ incorporates weak isospin and hypercharge and gives the corresponding fermion sectors the gauge quantum numbers of three SM generations under a common $\SUthreeC\times\SUtwoL\times\UoneY$ gauge algebra. This is the main representation-theoretic result of the construction.

The comparison with other three-generation mechanisms shows why the occurrence of the number three is not, by itself, sufficient to identify a common explanation of fermion families. In different approaches, a threefold structure may arise from representation multiplicities, triality, the rank-three structure of an exceptional Jordan algebra, nested division subalgebras, or a permutation symmetry imposed on three pre-existing fermion copies. These mechanisms address different aspects of the generation problem and need not have the same physical interpretation. The distinguishing feature of the present framework is the simultaneous construction of the fermion sectors, family action, and invariant gauge generators within a single Clifford algebra setting. This does not prove that the sedenions are uniquely selected by more fundamental principles, nor that nature must contain exactly three generations. It establishes instead that three gauge-equivalent fermion sectors follow from the stated algebraic assumptions.

The most immediate physical problem is to explain how the exact family symmetry is broken. In the present construction, the three sectors are treated symmetrically and are not intrinsically identified with the observed first, second, and third generations. A realistic extension must distinguish them dynamically and account for fermion mass hierarchies, the CKM and PMNS mixing matrices, and CP-violating phases. Subsequent work has begun to explore Higgs and flavour structures in this setting \cite{Gresnigt2026Higgs}, but these developments lie beyond the scope of the present review. The important question is whether the algebraic structure genuinely restricts the form of family-symmetry breaking and mixing, rather than merely providing a setting in which additional phenomenological parameters can be introduced.

A separate structural question concerns the relation between the sedenion-induced $S_3$ family action and $\Spin(8)$ triality. Although both involve $\Cl{8}$ and an $S_3$ symmetry, they are not the same action in the present formulation. The generator $\psi_3$ acts on the full Clifford algebra and does not preserve its canonical one-vector generating space; it therefore does not implement the standard outer triality action associated with $\Spin(8)$. It remains to determine whether the two $S_3$ structures can be related through a different choice of variables or a larger common algebraic setting, or whether they represent genuinely distinct uses of the same underlying mathematical structures.

A further challenge is the incorporation of spacetime symmetry and dynamics. The Clifford spinors considered here describe internal fermion structure and should not be identified automatically with Lorentz spinors. A complete physical theory must specify how the internal $\Cl{8}$ and $\Cl{10}$ constructions are combined with Lorentz symmetry, spacetime chirality, dynamical fields, and an action principle. Related work has explored how octonions and tensor products of division algebras can arise naturally within causal fermion systems, which in turn provide spacetime structures and dynamical equations through the causal action principle \cite{Finster2024Causal}. This does not yet constitute an embedding of the present $\Cl{8}$ and $\Cl{10}$ family construction into a causal fermion system, but it provides a relevant example of how division-algebraic internal structures may be connected to a dynamical spacetime framework. A complete extension must also explain the dynamics of the gauge and matter sectors rather than only their representation content. The present framework should therefore be regarded as a definite algebraic mechanism for organising three generations and their common gauge symmetries, not as a complete particle-physics theory. Its next important test is whether the same structure can constrain physics beyond representation theory and lead to genuinely predictive relations among the observed fermion generations.

\section*{Acknowledgements}
The author thanks Liam Gourlay for useful discussions and Janek Kozicki for independent Mathematica verification of the calculations.

\section*{Declaration of generative AI use}

During the preparation of this manuscript, the author used ChatGPT (OpenAI) to assist with language editing, organisation, the comparison of related approaches, and checks of notation, internal consistency, and bibliographic information. The tool was not used to establish the scientific results or conclusions of the article. All AI-assisted material was critically reviewed, checked against the cited literature where appropriate, and revised by the author, who takes full responsibility for the content of the manuscript.


\appendix

\section{Sedenion multiplication table}
\label{app:sedenion-multiplication-table}

For reference, Table~\ref{tab:sedenions} records the sedenion multiplication convention used throughout this review. The entry in the row labelled $s_i$ and the column labelled $s_j$ gives the product $s_is_j$; in particular, $s_0=1$ is the identity element.

\begin{table}[htbp]
\centering
\resizebox{0.98\textwidth}{!}{%
\rowcolors{2}{gray!10}{white}
\begin{tabular}{|c|*{16}{c|}}
\hline
\(\cdot\) & \(s_{0}\) & \(s_{1}\) & \(s_{2}\) & \(s_{3}\) & \(s_{4}\) & \(s_{5}\) & \(s_{6}\) & \(s_{7}\) 
          & \(s_{8}\) & \(s_{9}\) & \(s_{10}\) & \(s_{11}\) & \(s_{12}\) & \(s_{13}\) & \(s_{14}\) & \(s_{15}\)\\
\hline
\(s_0\)&
 \(s_{0}\)&\(s_{1}\)&\(s_{2}\)&\(s_{3}\)&\(s_{4}\)&\(s_{5}\)&\(s_{6}\)&\(s_{7}\)&
 \(s_{8}\)&\(s_{9}\)&\(s_{10}\)&\(s_{11}\)&\(s_{12}\)&\(s_{13}\)&\(s_{14}\)&\(s_{15}\)\\
\hline
\(s_1\)&
 \(s_{1}\)&\(-s_{0}\)&\(s_{3}\)&\(-s_{2}\)&\(s_{5}\)&\(-s_{4}\)&\(-s_{7}\)&\(s_{6}\)&
 \(s_{9}\)&\(-s_{8}\)&\(-s_{11}\)&\(s_{10}\)&\(-s_{13}\)&\(s_{12}\)&\(s_{15}\)&\(-s_{14}\)\\
\hline
\(s_2\)&
 \(s_{2}\)&\(-s_{3}\)&\(-s_{0}\)&\(s_{1}\)&\(s_{6}\)&\(s_{7}\)&\(-s_{4}\)&\(-s_{5}\)&
 \(s_{10}\)&\(s_{11}\)&\(-s_{8}\)&\(-s_{9}\)&\(-s_{14}\)&\(-s_{15}\)&\(s_{12}\)&\(s_{13}\)\\
\hline
\(s_3\)&
 \(s_{3}\)&\(s_{2}\)&\(-s_{1}\)&\(-s_{0}\)&\(s_{7}\)&\(-s_{6}\)&\(s_{5}\)&\(-s_{4}\)&
 \(s_{11}\)&\(-s_{10}\)&\(s_{9}\)&\(-s_{8}\)&\(-s_{15}\)&\(s_{14}\)&\(-s_{13}\)&\(s_{12}\)\\
\hline
\(s_4\)&
 \(s_{4}\)&\(-s_{5}\)&\(-s_{6}\)&\(-s_{7}\)&\(-s_{0}\)&\(s_{1}\)&\(s_{2}\)&\(s_{3}\)&
 \(s_{12}\)&\(s_{13}\)&\(s_{14}\)&\(s_{15}\)&\(-s_{8}\)&\(-s_{9}\)&\(-s_{10}\)&\(-s_{11}\)\\
\hline
\(s_5\)&
 \(s_{5}\)&\(s_{4}\)&\(-s_{7}\)&\(s_{6}\)&\(-s_{1}\)&\(-s_{0}\)&\(-s_{3}\)&\(s_{2}\)&
 \(s_{13}\)&\(-s_{12}\)&\(s_{15}\)&\(-s_{14}\)&\(s_{9}\)&\(-s_{8}\)&\(s_{11}\)&\(-s_{10}\)\\
\hline
\(s_6\)&
 \(s_{6}\)&\(s_{7}\)&\(s_{4}\)&\(-s_{5}\)&\(-s_{2}\)&\(s_{3}\)&\(-s_{0}\)&\(-s_{1}\)&
 \(s_{14}\)&\(-s_{15}\)&\(-s_{12}\)&\(s_{13}\)&\(s_{10}\)&\(-s_{11}\)&\(-s_{8}\)&\(s_{9}\)\\
\hline
\(s_7\)&
 \(s_{7}\)&\(-s_{6}\)&\(s_{5}\)&\(s_{4}\)&\(-s_{3}\)&\(-s_{2}\)&\(s_{1}\)&\(-s_{0}\)&
 \(s_{15}\)&\(s_{14}\)&\(-s_{13}\)&\(-s_{12}\)&\(s_{11}\)&\(s_{10}\)&\(-s_{9}\)&\(-s_{8}\)\\
\hline
\(s_8\)&
 \(s_{8}\)&\(-s_{9}\)&\(-s_{10}\)&\(-s_{11}\)&\(-s_{12}\)&\(-s_{13}\)&\(-s_{14}\)&\(-s_{15}\)&
 \(-s_{0}\)&\(s_{1}\)&\(s_{2}\)&\(s_{3}\)&\(s_{4}\)&\(s_{5}\)&\(s_{6}\)&\(s_{7}\)\\
\hline
\(s_9\)&
 \(s_{9}\)&\(s_{8}\)&\(-s_{11}\)&\(s_{10}\)&\(-s_{13}\)&\(s_{12}\)&\(-s_{15}\)&\(s_{14}\)&
 \(s_{1}\)&\(-s_{0}\)&\(-s_{3}\)&\(s_{2}\)&\(-s_{5}\)&\(s_{4}\)&\(-s_{7}\)&\(s_{6}\)\\
\hline
\(s_{10}\)&
 \(s_{10}\)&\(s_{11}\)&\(s_{8}\)&\(-s_{9}\)&\(-s_{14}\)&\(-s_{15}\)&\(s_{12}\)&\(s_{13}\)&
 \(s_{2}\)&\(s_{3}\)&\(-s_{0}\)&\(-s_{1}\)&\(-s_{6}\)&\(-s_{7}\)&\(s_{4}\)&\(s_{5}\)\\
\hline
\(s_{11}\)&
 \(s_{11}\)&\(-s_{10}\)&\(s_{9}\)&\(s_{8}\)&\(-s_{15}\)&\(s_{14}\)&\(-s_{13}\)&\(s_{12}\)&
 \(s_{3}\)&\(-s_{2}\)&\(s_{1}\)&\(-s_{0}\)&\(-s_{7}\)&\(s_{6}\)&\(-s_{5}\)&\(s_{4}\)\\
\hline
\(s_{12}\)&
 \(s_{12}\)&\(s_{13}\)&\(s_{14}\)&\(s_{15}\)&\(s_{8}\)&\(-s_{9}\)&\(-s_{10}\)&\(-s_{11}\)&
 \(s_{4}\)&\(s_{5}\)&\(s_{6}\)&\(s_{7}\)&\(-s_{0}\)&\(-s_{1}\)&\(-s_{2}\)&\(-s_{3}\)\\
\hline
\(s_{13}\)&
 \(s_{13}\)&\(-s_{12}\)&\(s_{15}\)&\(-s_{14}\)&\(s_{9}\)&\(s_{8}\)&\(s_{11}\)&\(-s_{10}\)&
 \(s_{5}\)&\(-s_{4}\)&\(s_{7}\)&\(-s_{6}\)&\(s_{1}\)&\(-s_{0}\)&\(s_{3}\)&\(-s_{2}\)\\
\hline
\(s_{14}\)&
 \(s_{14}\)&\(-s_{15}\)&\(-s_{12}\)&\(s_{13}\)&\(s_{10}\)&\(-s_{11}\)&\(s_{8}\)&\(s_{9}\)&
 \(s_{6}\)&\(-s_{7}\)&\(-s_{4}\)&\(s_{5}\)&\(s_{2}\)&\(-s_{3}\)&\(-s_{0}\)&\(s_{1}\)\\
\hline
\(s_{15}\)&
 \(s_{15}\)&\(s_{14}\)&\(-s_{13}\)&\(-s_{12}\)&\(s_{11}\)&\(s_{10}\)&\(-s_{9}\)&\(s_{8}\)&
 \(s_{7}\)&\(s_{6}\)&\(-s_{5}\)&\(-s_{4}\)&\(s_{3}\)&\(s_{2}\)&\(-s_{1}\)&\(-s_{0}\)\\
\hline
\end{tabular}
}
\caption{Sedenion multiplication table using \(s_i\).}
\label{tab:sedenions}
\end{table}

\bibliographystyle{unsrt}
\bibliography{references_cleaned}


\end{document}